\documentclass[letterpaper,twocolumn,10pt]{article}
\usepackage{include/usenix}

\usepackage{tikz}
\usepackage{float}
\usepackage{amsmath}
\usepackage{balance}
\usepackage{booktabs}
\usepackage{hyperref}
\usepackage[capitalise,nameinlink]{cleveref}
\usepackage{xcolor}
\usepackage{xspace}
\usepackage{multirow}
\usepackage{tabularx}
\usepackage{algorithm}
\usepackage{algpseudocode}
\algtext*{EndWhile}
\algtext*{EndIf}
\algtext*{EndFor}
\usepackage{caption}
\usepackage{subcaption}
\usepackage[compact]{titlesec}

\usepackage[inline]{enumitem}

\usepackage{enumitem}
\setitemize{noitemsep,topsep=0pt,parsep=0pt,partopsep=0pt}

\newcommand{\sys}[0]{CRIUgpu\xspace}
\def\Snospace~{\S{}}

\newcommand{\stitle}[1]{\vspace{1.ex}\noindent{\bf #1}}

\usepackage[compact]{titlesec}
\titleformat*{\section}{\large\bfseries}
\titleformat*{\subsection}{\normalsize\bfseries}
\titleformat*{\subsubsection}{\normalsize\bfseries}
\titlespacing*{\section}{0pt}{*2}{*1}
\titlespacing*{\subsection}{0pt}{*1.5}{*0.8}

\usepackage{tikz}
\newcommand*\circled[1]{
    \tikz[baseline=(char.base)]{
        \node[shape=circle,draw,inner sep=.5pt] (char) {#1};
    }
}

\newcommand{\rulesep}{\unskip\ \vrule\ }

\begin{document}


\title{\sys: Transparent Checkpointing of GPU-Accelerated Workloads}

\author{
    \rm
        Radostin Stoyanov$^{\star\dagger}$ \enskip
        Viktória Spišaková$^{\ddagger}$ \enskip
        Jesus Ramos$^{\diamond}$ \enskip
        Steven Gurfinkel$^{\diamond}$ \enskip
        Andrei Vagin$^{\triangle}$ \enskip
    \\ \rm
        Adrian Reber$^{\dagger}$ \enskip
        Wesley Armour$^{\star}$ \enskip
        Rodrigo Bruno$^{\bullet}$ \enskip
    \\ \\
    {
        $^{\star}$University of Oxford\enskip
        {$^{\ddagger}$Masaryk University\enskip}
        $^{\diamond}$NVIDIA\enskip 
        $^{\triangle}$Google\enskip
        $^{\dagger}$Red Hat\enskip
    }\\
        {$^{\bullet}$INESC-ID, Instituto Superior Técnico, University of Lisbon\enskip}
}

\maketitle

\begin{abstract}
Deep learning training at scale is resource-intensive and time-consuming, often running across hundreds or thousands of GPUs for weeks or months. Efficient checkpointing is crucial for running these workloads, especially in multi-tenant environments where compute resources are shared, and job preemptions or interruptions are common. However, transparent and unified GPU snapshots are particularly challenging because of the hardware architecture differences between CPU and GPU, including memory subsystems, dynamic parallelism, and thread synchronization. State-of-the-art GPU checkpointing techniques typically leverage mechanisms that intercept, log, and replay device API calls. However, this approach adds performance overhead and requires hardware-specific implementation that is difficult to test, maintain, and integrate with existing container platforms. In this paper, we present \sys~-- a novel approach for transparent checkpointing of GPU-accelerated workloads that builds on recently introduced driver capabilities, enabling support for CUDA and ROCm applications. Our evaluation results show that \sys works with a variety of deep learning and high-performance computing workloads running across multiple GPUs, completely eliminating steady-state performance overheads, and significantly reducing recovery times compared to state-of-the-art transparent GPU checkpointing mechanisms.
\end{abstract}
\section{Introduction}%
In recent years, advances in high-performance computing (HPC) and machine learning have enabled exponential growth in neural network sizes with increasing number of parameters and layers, unlocking a wide range of applications, from conversational bots like OpenAI's ChatGPT~\cite{chatgpt}, Google's Gemini~\cite{team2024gemini}, and Microsoft's Copilot~\cite{copilot} to programming assistants~\cite{chen2021evaluating, microsoft2023github, roziere2024code}, and autonomous driving vehicles~\cite{kiran2022deep,ma2023dolphins}. However, training large models at an unprecedented scale presents new and interesting system challenges. A recent study has shown that training LLaMA 3.1 model with 405 billion parameters on 16,000 GPUs has encountered 419 unexpected interruptions over 54 days, with 78\% attributed to hardware failures~\cite{dubey2024llama}.
These errors, as demonstrated in~\cite{gupta2024just}, can cost up to few million dollars per month, depending on the size of the job. To handle such disruptions, training large models such as Google Gemini often requires maintaining redundant in-memory model states to enable rapid recovery~\cite{team2024gemini}.


\begin{figure}[t]
    \centering
    \includegraphics[width=\columnwidth]{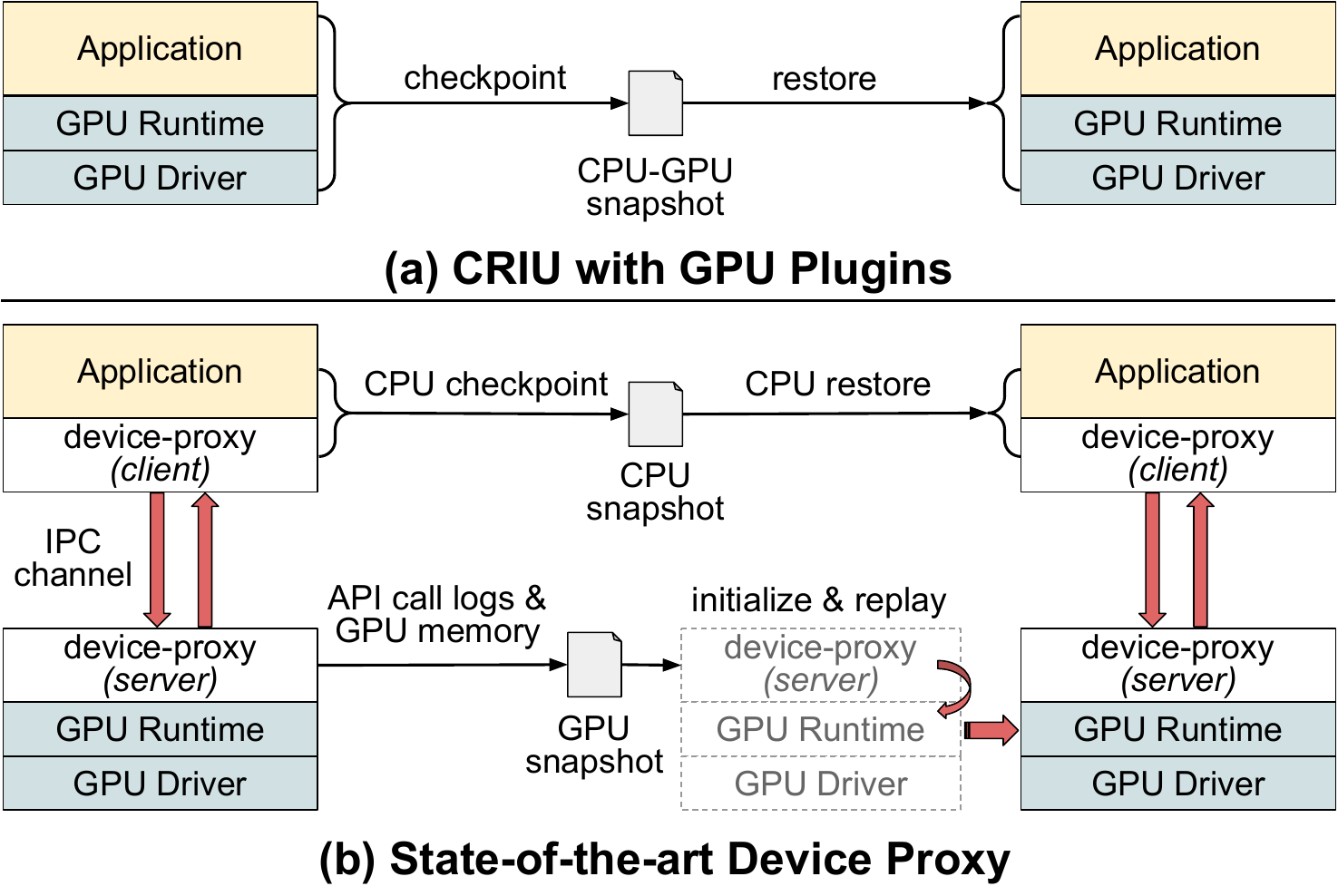}
    \caption{A comparison between (a) \sys and (b) state-of-the-art checkpointing system using a mechanism for interception, logging and replay of device API calls.}
    \label{fig:cuda-device-proxy}
\end{figure}

GPU checkpointing provides an effective failure recovery solution that allows to periodically save the state of models and resume training from the last checkpoint. This checkpointing mechanism can be implemented at different levels of the software stack: \textit{application-level} (not transparent), \textit{library-level} (semi-transparent), and \textit{system-level} (fully transparent). System-level checkpointing makes no assumptions about the underlying tasks or processes, offering a more \textit{general} and \textit{transparent} solution that is easy to use and does not require changes to application code~\cite{chaudhary2020balancing,shukla2022singularity,gupta2024just}. However, enabling transparent GPU checkpointing at system-level is particularly challenging, as it requires specialized knowledge of the hardware architecture, as well as an understanding of the interconnected memory subsystems, dynamic parallelism, and thread synchronization. These low-level implementation details often change significantly across hardware generations and are not always publicly documented. As a result, state-of-the-art GPU checkpointing solutions rely on \textit{device proxy} mechanisms that decouple the memory address space of applications from the device-specific mappings and other side-effects created by GPU libraries such as CUDA~\cite{jain2020crac, shukla2022singularity,eiling2023checkpoint}. This approach dynamically loads a shared library with the \texttt{LD\_PRELOAD} mechanism that allows to intercept, log, and replay all device APIs invoked by the application. However, this interception mechanism is in the critical path of GPU-accelerated operations and inherently introduces steady-state performance overhead with each API call, as shown in~\Cref{sec:background}. In addition, it adds significant complexity and maintenance overhead, as it requires low-latency communication between the host processes and the device-proxy component, custom memory management, synchronization barriers for distributed jobs, and explicit handling of each API call. It also creates additional runtime dependencies and potential compatibility issues, as it requires recompiling workloads such as PyTorch from source with dynamically linked libraries (e.g., CUDA runtime)~\cite{eiling2023cricket}. Replaying the logged API calls during restore can also lead to prolonged recovery times and inconsistent GPU state, especially with non-deterministic operations such as floating-point computations~\cite{riachnvidia2019gtc}.

To address these challenges, we propose a novel approach for transparent checkpoint/restore of GPU-accelerated workloads that builds on recently introduced GPU driver checkpointing capabilities~\cite{gurfinkel2024checkpointing, bhardwaj2021fast}. This approach does not require interception of device API calls, and enables \textit{unified} system-level snapshots that contain both CPU and GPU state, as illustrated in \Cref{fig:cuda-device-proxy}. We implement the proposed checkpointing functionality on top of the open-source Checkpoint/Restore in Userspace (CRIU)~\cite{criu} tool through plugins that handle all external GPU resources. We extensively study the checkpoint/restore performance of \sys with large language model training for BERT, GPT-2, and LLaMA 3. We further evaluate \sys on various hardware configurations, including single GPU and multi-GPU setups, using NVIDIA H100, A100, V100, and A6000. We also demonstrate \sys's capability to efficiently checkpoint and restore HPC applications running on AMD GPU devices. 

Our evaluation results show that \sys significantly reduces checkpoint and restore overheads in comparison to state-of-the-art API interception mechanisms, while enabling reliable checkpoint/restore of GPU-accelerated containers.
To the best of our knowledge, \sys is the first to achieve fully transparent, system-level GPU container checkpoints that combine both GPU and CPU state in a single unified snapshot. We have contributed the changes implementing \sys to the upstream GitHub repository, which were released with CRIU version 4.0 and are publicly available at \url{https://github.com/checkpoint-restore/criu}.

In summary, this paper makes the following contributions:

\begin{itemize}
    \item We systematically analyze state-of-the-art GPU checkpoint mechanisms utilizing device API call interception techniques and uncover several challenges that limit the usability of this approach. We further benchmarked the only open-source system we could find that implements these mechanisms (Cricket~\cite{eiling2022cricket}) and measured the performance overheads when compared to a baseline neural network training with no API interception (\textsection{\ref{sec:background}});
    
    \item We propose \sys, a novel approach for fully transparent and unified checkpointing of GPU-accelerated workloads, including deep learning training and inference. \sys eliminates steady-state overheads, supports both AMD and NVIDIA GPUs, and, to the best of our knowledge, is the first to support fully transparent checkpointing of large multi-GPU applications without device API call interception (\textsection{\ref{sec:design}});

    \item Our prototype builds on the open-source CRIU project integrated with Podman (a widely used container engine) to enable transparent and unified checkpointing of GPU containers (\textsection{\ref{sec:implementation}}). We have contributed our implementation to the corresponding open-source projects.

    \item We evaluate {\sys} on large language model training workloads, such as GPT-2 (1.5B) and LLaMA 3.1 (8B), as well as a set of HPC micro-benchmarks, showing that {\sys} can be used to checkpoint and restore such models on different GPU devices (e.g., NVIDIA H100, A100, and AMD MI210) with minimal overhead (\textsection{\ref{sec:evaluation}}).
\end{itemize}

\begin{figure*}[t]
    \centering
    \begin{subfigure}[t]{0.49\textwidth}
        \centering
        \includegraphics[width=\textwidth]{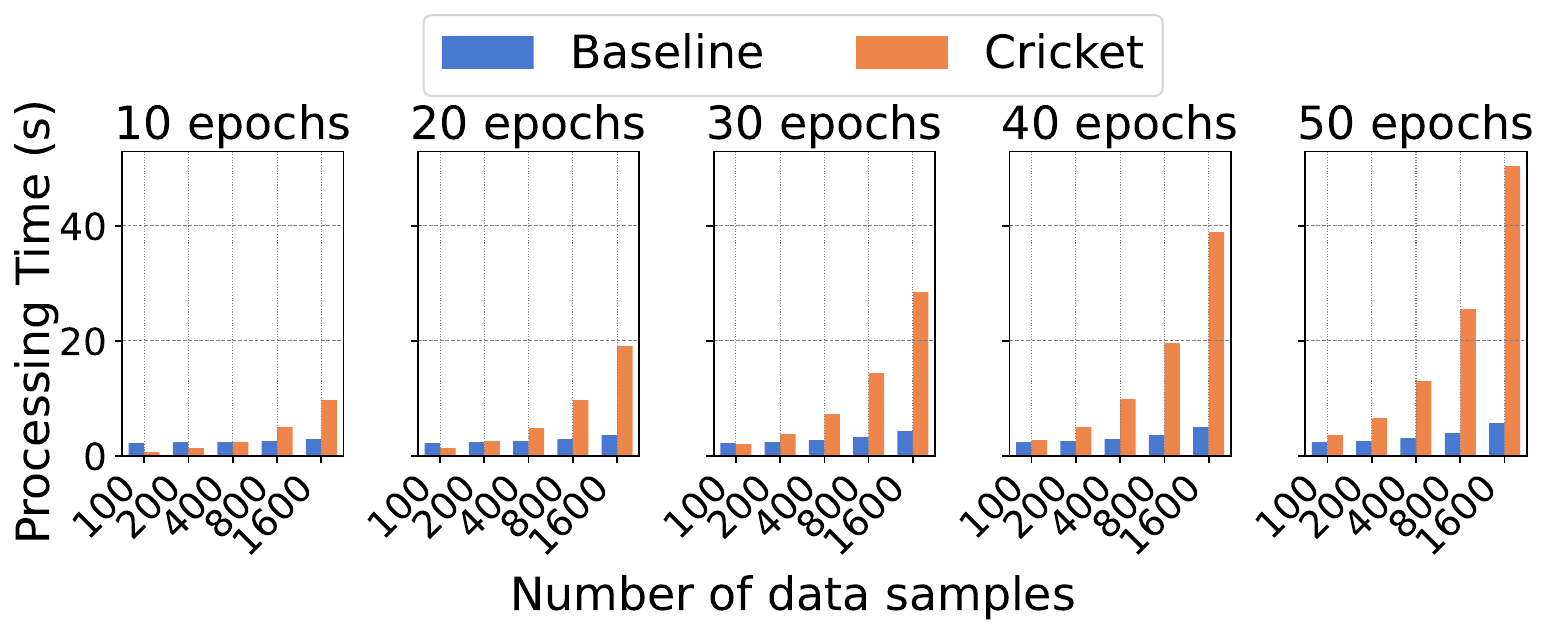}
    \end{subfigure}
    \hfill
    \begin{subfigure}[t]{0.5\textwidth}
        \centering
        \includegraphics[width=\textwidth]{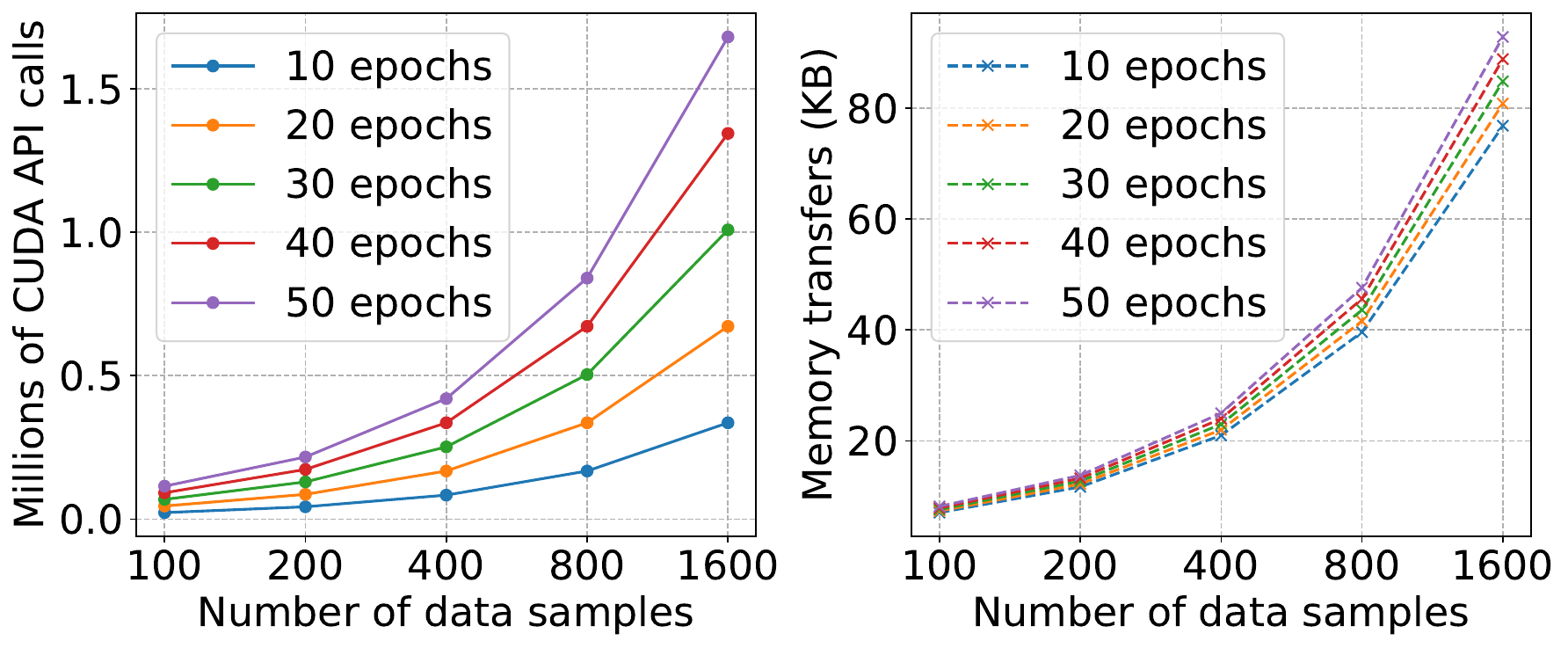}
    \end{subfigure}
    \vspace{-.5em}
    \caption{Analysis of intercepted CUDA API calls and memory transfers between host and device during neural network training reveals significant overhead that increases with the number of epochs.~\textbf{\textup{Setup}}: A PyTorch implementation of stochastic gradient descent neural network training with one input layer (10 features), one hidden layer (50 units), and one output layer (1 unit) running \textit{with (Cricket)} and \textit{without (Baseline)} API interception.}
    \label{fig:cricket-vs-baseline}
    \vspace{-1em}
\end{figure*}
\section{Background and Motivation}\label{sec:background}%

Deep learning training consumes significant computational resources, including processing power and memory. For example, training a state-of-the-art large language model (LLM) such as LLaMA 3.1 with 405 billion parameters requires a cumulative 39.3 million hours of GPU computation on H100 80GB device~\cite{dubey2024llama}. The training process typically starts with a random set of learnable parameters and processes a small disjoint subset (\textit{minibatch}) of a training data set in steps known as \textit{iterations}. When the entire dataset is processed exactly once, an \textit{epoch} is said to be complete. Each training iteration involves data augmentation, a forward pass to generate predictions, a backward pass to compute gradients, and a weights update phase~\cite{goodfellow2016deep}. Once training is completed (typically after a predetermined number of epochs), the final learned parameters are saved as a checkpoint file to persistent storage, enabling the model to be initialized from this file for \textit{inference} to generate predictions.

To accelerate model training, traditional cloud platforms are increasingly leveraging GPUs to enable parallel and distributed computations. However, efficiently managing cluster resources with tens of thousands of GPUs across multiple teams is challenging, especially since these workloads often run for weeks or months and consume enormous amounts of energy~\cite{yang2024part, dubey2024llama}. When running at scale, hardware faults, network interruptions, and software bugs occur frequently, and each individual fault can result in partial restarts or a complete retraining from scratch~\cite{rojas2019analyzing}. As a result, efficient fault tolerance and checkpointing techniques have become of paramount importance for large-scale model training~\cite{mohan2021checkfreq,wang2023gemini,gupta2024just}.

\subsection{Challenges with Transparent GPU Checkpointing}%
Several approaches for transparent system-level checkpointing of GPU applications have been proposed in the literature~\cite{takizawa2009checuda,nukada2011nvcr,jain2020crac,shukla2022singularity,eiling2022cricket,eiling2023checkpoint}. However, these approaches are not \textit{fully transparent} as they require invasive modifications to the execution environment of applications to intercept, log, and replay device API calls, i.e., \textit{semi-transparent} checkpointing. \Cref{fig:cuda-device-proxy} illustrates this mechanism. To enable GPU checkpointing, these solutions assume that all interactions with the GPU go through specific dynamically linked libraries (e.g., \texttt{CUDA} for NVIDIA GPUs or \texttt{ROCm} for AMD GPUs)~\cite{shukla2022singularity,eiling2022cricket}. To enable checkpointing, the function signatures provided by these libraries are overridden by an interception library, which is preloaded using the \texttt{LD\_PRELOAD} environment variable. However, implementing this interception mechanism is challenging due to the complex architecture of GPU devices, which can vary significantly across hardware generations, and the proprietary nature of the GPUs libraries (e.g., NVML, cuDNN). We outline some of the key challenges of transparent GPU checkpointing below.

\subsubsection*{Challenge 1: API Interception with a Device Proxy}%
CUDA offers multiple ways for applications to interface the GPU driver -- a \textit{high-level} (runtime)~\cite{nvidia2024runtime} and \textit{low-level} (driver)~\cite{nvidia2024driver} APIs. The runtime API simplifies application development by offering implicit initialization, context management, and module handling. In contrast, the driver API provides fine-grained control with explicit context and module management, offering greater flexibility but requiring a more complex implementation.
To maximize checkpoint transparency, API calls should be intercepted at a low level -- ideally, between the driver API and the GPU driver~\cite{shukla2022singularity}. However, due to the closed-source nature of GPU drivers, some checkpointing solutions intercept only the runtime API~\cite{eiling2022cricket}. Since all interactions between the CPU and GPU must pass through the interception mechanism, scenarios where host memory addresses are used as implicit input and accessed directly by the GPU could potentially lead to an inconsistent state after restore~\cite{gupta2024just}.
In addition, to enable the replay of all device operations, a worker process must log these device APIs, along with their input values, object handles (e.g., events, streams), GPU memory addresses, and input parameters. These interception and logging operations consume additional CPU and memory resources, inevitably introducing performance overhead and increasing API call latency~\cite{eiling2022cricket}.
As \sys does not require an API interception mechanism, it avoids these challeges completely.

\subsubsection*{Challenge 2: Static and Dynamic Linking}%
Starting with CUDA version 5.5, the CUDA runtime library is statically linked by default when using the CUDA compiler~\cite{corporation2023cuda}. This static linking ensures that the correct version of runtime functions are included in the application binary and avoid the need for additional redistribution of CUDA libraries. This default static linking is widely used by frameworks such as PyTorch to enable easier cross-platform application development.
However, a prerequisite for semi-transparent checkpointing is for the CUDA runtime library to be dynamically-linked to enable the replacement with interception code~\cite{eiling2022cricket}.
As a result, many applications and frameworks need to be modified and recompiled from source to meet this requirement, which may involve significant effort and be impractical when the source code is unavailable. In contrast, \sys interacts directly with the GPU driver, allowing to checkpoint applications that use both static and dynamic linking.

\subsubsection*{Challenge 3: Starting kernels from Shared Objects}%
For some applications, such as PyTorch, CUDA kernels are loaded at runtime using \texttt{dlopen} and \texttt{dlsym}, with the initialization phase enabling the invocation of \texttt{cuLaunchKernel()}. However, when using semi-transparent checkpointing, the CPU and GPU states are decoupled, requiring these kernels to be dynamically loaded in both address spaces. This is typically achieved by intercepting CUDA-related registration functions and sending a request to dynamically load the CUDA kernels into the address space of the interception (device-proxy) server. Implementing this functionality requires reverse-engineering part of the CUDA runtime to decode the metadata of fat binaries (which store multiple versions of binary code for different GPU architectures~\cite{harris2024cuda}), extracting the corresponding cubin from the target binary, and sending it through the interception mechanism to be loaded with \texttt{cuModuleLoadData}. This results in additional complexity and performance overhead. In comparison, \sys avoids this challenge by interacting with the GPU driver directly and utilizing its functionality lock/unlock the execution of APIs that impact the GPU state.  

\subsubsection*{Challenge 4: Complex GPU Runtime and Memory State}%
GPUs have a complex hierarchy of multi-processors, processing blocks, and memory controllers that facilitate high degree of parallel computations and memory bandwidth to maximize compute performance~\cite{nvidia2022h100, nvidia2020a100, nvidia2017p100}.
GPU code execution breaks down into three stages: (i) launch, (ii) grid initialization, and (iii) kernel execution. Many GPU-intensive applications, such as deep neural network training and scientific simulations, involve iterative processes where the same workflow is executed repeatedly. In these cases, CUDA streams -- which allow for concurrent execution of tasks on the GPU -- require the CPU to resubmit the same work to the GPU after each iteration. This resubmission is necessary to ensure that the GPU can continue processing each task in parallel, but it can introduce overhead in applications with many iterations. An alternative, more efficient, method for submitting work to the GPU is using \textit{task graphs} that consist of a series of operations such as memory copies and kernel launches connected by dependencies. This method enables allows reduce the cost of kernel launch significantly. However, the inherent complexity and advanced optimizations of the GPU runtime used to manage this state further complicates the API interception mechanisms for semi-transparent checkpointing.

\subsubsection*{Challenge 5: No Internal GPU State Access}%
In contrast to CPUs, most GPUs do not offer an assembler that maps the assembly code directly to machine instructions with register access. This limitation poses a significant challenge to implementing transparent checkpoint/restore operations because some register state can only be accessed using assembly. For instance, Cricket~\cite{eiling2022cricket} addresses this problem using Streaming ASSembly (SASS) by directly modifying the binary code after compilation to recreate the device execution state. Singularity~\cite{sivathanu2022transparent} uses similar approach where the \texttt{cuObjDump} utility is used to parse kernel libraries and extract parameter information, then intercept \texttt{nvrtcCompileProgram} to extract the parameter signatures from the generated parallel-thread execution (PTX) program. However, both solutions have high performance cost, are prone to errors due to the complexity of reverse-engineering, and may not be universally applicable across different GPU generations. \sys avoids these limitations by utilizing the driver directly to restore the GPU state from a checkpoint. 

\subsubsection*{Challenge 6: Determinism and Reproducibility}%
Replaying CPU-GPU interactions correctly during restore is particularly challenging when applications use asynchronous or non-deterministic operations~\cite{park2022gpureplay}.
These operations can introduce state divergences when replaying the recorded API calls during restore and make it difficult to reproduce the same results across different runs.
This problem is further exacerbated with long-running workloads, such as deep learning training jobs with parallel computations running across multiple GPUs, where small state discrepancies accumulate over time~\cite{riachnvidia2019gtc}. For example, frameworks such as Megatron-LM are intended to be bitwise reproducible~\cite{nvidia2024megatron-lm}, where the same training application run twice in the same hardware and software environment should produce identical checkpoints, losses, and accuracy values.
Existing work aims to address these challenges by removing the sources of non-determinism at the record time (e.g., by forcing synchronous operations) and tolerating non-deterministic interactions that do not affect the GPU state at the replay time~\cite{park2022gpureplay,gupta2024just}.
In contrast, \sys does not use record-and-replay of CPU-GPU interactions and relies on a locking mechanism to create snapshots and restore them in a consistent and deterministic state.

\subsection{Performance Overhead of API Interception}%
To better understand the overhead introduced by semi-transparent checkpointing with API call interception, we analyze the performance of Cricket~\cite{eiling2022cricket}, a state-of-the-art open-source GPU checkpointing library.
By examining Cricket's implementation, we gain insights into the typical costs associated with the interception layer and saving and restoring the GPU state.
As part of our analysis, we conduct an experiment using a simple PyTorch implementation of neural network training, where we measure the number of intercepted API calls and the total processing time to quantify the impact of the interception layer on overall performance.
The results in \Cref{fig:cricket-vs-baseline} show that the number of intercepted API calls and the overhead introduced by the interception layer increase exponentially with the number of training epochs.
In addition, the handling of asynchronous data transfers is achieved at the interception layer by simply forwarding \texttt{cudaMemcpyAsync} to \texttt{cudaMemcpy} and ignoring the stream parameter~\cite{eiling2022cricket}.

\stitle{Summary.} Compared to semi-transparent checkpointing, \sys does not require interception mechanisms, re-complication of the application code, and supports checkpointing with both statically linked and dynamically linked libraries. \sys relies on the GPU driver to save and restore the state of applications.
In particular, it leverages recently introduced checkpoint/restore driver capabilities and combines these with CPU-checkpointing techniques to efficiently handle serialization and de-serialization of runtime state, memory and kernel parameters in both GPU and CPU.

\begin{figure*}[t]
    \centering
    \begin{subfigure}[b]{0.49\textwidth}
        \centering
        \includegraphics[width=\textwidth]{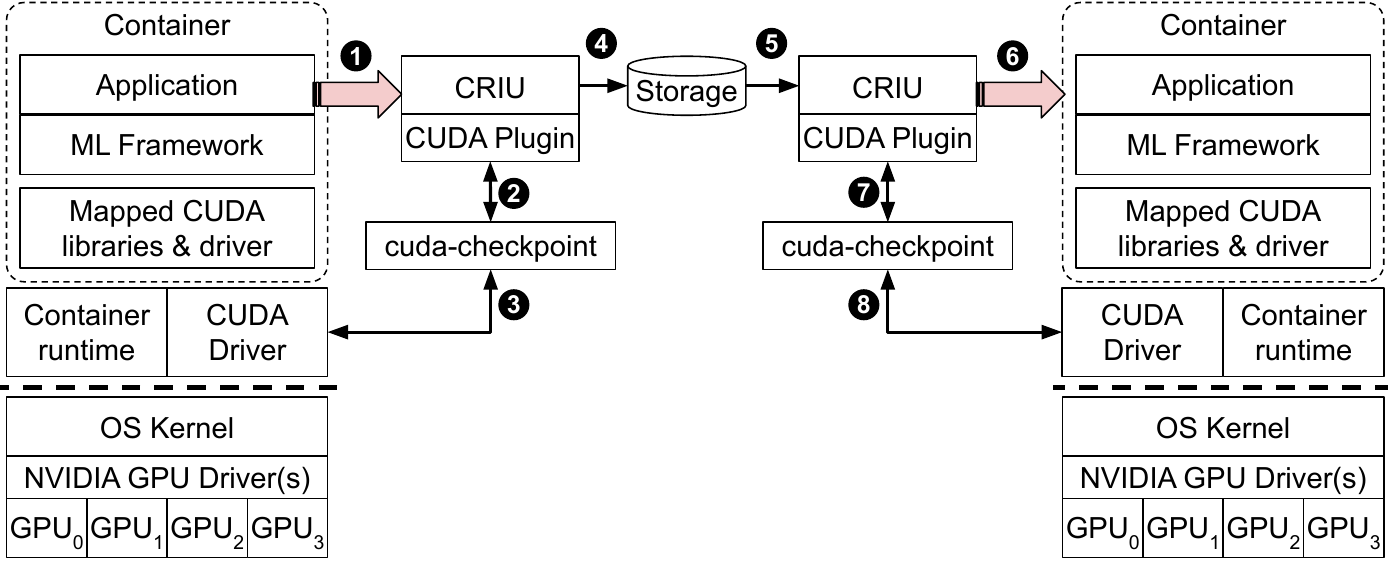}
        \caption{Checkpoint/restore with CUDA plugin.}
        \label{fig:criu-cuda-checkpoint-arch}
    \end{subfigure}
    \hspace{0.005em}
    \rulesep
    \hspace{0.005em}
    \begin{subfigure}[b]{0.49\textwidth}
        \centering
        \includegraphics[width=\textwidth]{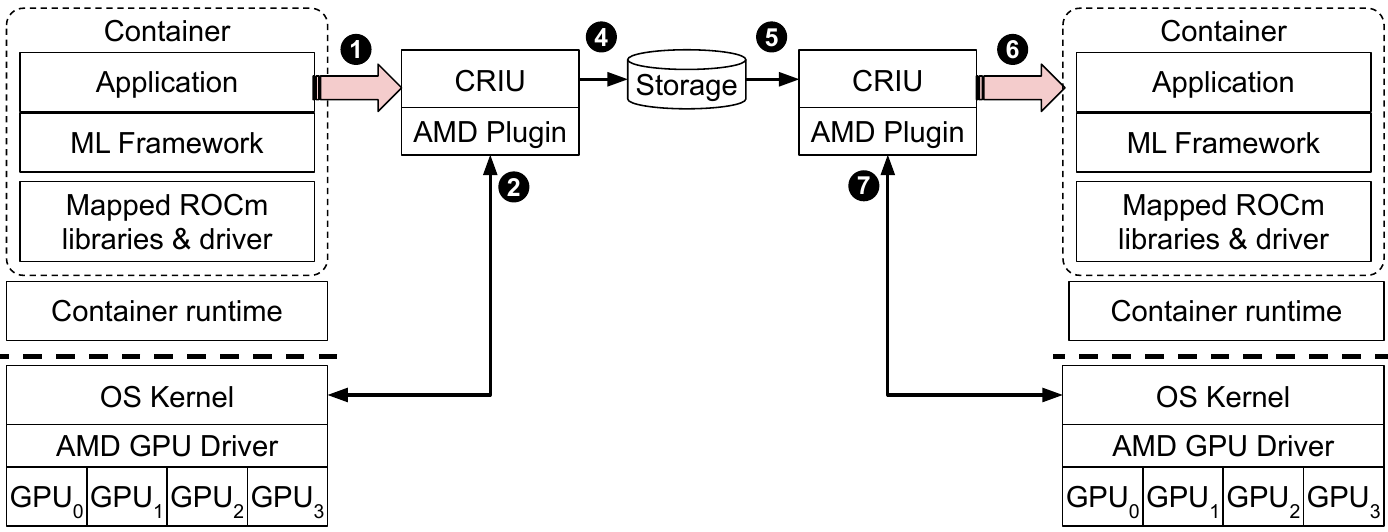}
        \caption{Checkpoint/restore with AMD GPU plugin.}
        \label{fig:criu-amdgpu-checkpoint-arch}
    \end{subfigure}
    \vspace{-0.5em}
    \caption{An overview of the transparent checkpoint/restore mechanisms with CUDA and AMD GPU plugins for CRIU.}
    \label{fig:criu-gpu-checkpoint-arch}
    \vspace{-1em}
\end{figure*}
\section{Transparent GPU Checkpointing}\label{sec:design}%
Several open-source tools enable transparent checkpointing of Linux processes running on the CPU~\cite{hhargrove2006berkeley,ansel2009dmtcp,criu}, of which Checkpoint/Restore in Userspace (CRIU) is the most widely used and actively maintained. However, a key limitation of CRIU is that, out of the box, it does not support saving and restoring the state of external hardware devices such as GPUs~\cite{shukla2022singularity,eiling2022cricket}. To address this limitation, we extend the functionality of CRIU with \textit{plugins} (\autoref{sec:gpu-plugins}) that handle GPU state. In comparison to the previous work (utilizing device-proxy mechanisms to intercept, log, and replay API calls~\cite{eiling2022cricket,eiling2023cricket,shukla2022singularity,gupta2024just}), we leverage recently introduced driver capabilities to enable transparent GPU checkpointing~\cite{bhardwaj2021drm,gurfinkel2024checkpointing}. Our aim is to enable \textit{fully transparent} checkpointing that supports a wide range of GPU devices and avoids the performance overheads and limitations of API interception (\autoref{sec:background}).

\subsection{GPU Plugins}\label{sec:gpu-plugins}
Checkpointing of CUDA~\cite{gurfinkel2024checkpointing} and ROCm~\cite{bhardwaj2021fast} applications is achieved through driver capabilities that capture and restore the GPU state (e.g., memory) associated with the target processes. Since this functionality is specific to GPU-accelerated applications and not required for other (e.g., CPU-only) workloads, we implement it as dynamically loadable shared libraries (\textit{plugins}), which can be optionally installed. When these plugins are installed, they are loaded during CRIU's initialization phase and utilized to handle GPU resources. ~\Cref{fig:criu-gpu-checkpoint-arch} illustrates the checkpoint/restore mechanisms with CUDA (\autoref{sec:cuda-plugin}) and AMD GPU (\autoref{sec:amd-gpu-plugin}) plugins. These plugins implement callbacks that are executed at specific stages (known as \textit{hooks}; \autoref{sec:plugin-hooks}) during the checkpoint and restore operations. In addition, each plugin defines \textit{initialization} and \textit{exit} callback functions. The initialization function is called when the plugin is loaded, with an argument specifying the current CRIU operation (\textit{dump}, \textit{pre-dump}, or \textit{restore}). Similarly, the plugin's exit function is invoked at the end of the checkpoint/restore operation, with an argument indicating whether the operation has been successful. This allows the plugins to perform cleanup tasks or, in the event of an error, restore the target processes to their original state.

\subsubsection{CUDA Plugin}\label{sec:cuda-plugin}%
The CUDA plugin utilizes a checkpointing utility called \texttt{cuda-checkpoint}~\cite{cuda-checkpoint} to perform a set of actions (\textit{lock}, \textit{checkpoint}, \textit{restore}, \textit{unlock}) for all tasks running on NVIDIA GPUs.
In particular, these actions are used to enable transparent GPU checkpointing as follows:
\begin{enumerate}[label=\itshape(\roman*\upshape),nosep]
    \item \textit{Locking} all CUDA APIs affecting the GPU state of the target processes and waiting for active operations (e.g., stream callbacks) to complete. \sys uses a timeout (10 seconds by default) with this action to avoid indefinite blocking. If the timeout expires, \sys attempts to restore all CPU and GPU tasks to their original state.

    \item \textit{Checkpointing} the GPU state of CUDA tasks into host memory allocations managed by the driver, and releasing all GPU resources held by the application.
\end{enumerate}
Executing these steps results in the CUDA tasks entering a \textit{checkpointed} state without direct reference to GPU hardware.
This allows to perform checkpoint/restore operations with CRIU similar to a CPU-only workloads.
It is important to note that a standalone invocation of the \texttt{cuda-checkpoint} tool does not handle the state of processes and threads running on the CPU, which can result in undefined behavior.
For example, multi-threaded workloads, such as Ollama~\cite{morgan2023ollama}, use error-handling mechanisms that detect unresponsive GPU tasks and restart them.
To prevent inconsistencies and undefined behavior, \sys ensures that all CPU and GPU tasks are suspended (locked) before checkpointing their state.
This is achieved through the Linux ptrace seize with interrupt mechanism~\cite{linux-ptrace}, which halts the execution of relevant processes and threads running on the CPU, allowing \sys to capture their state in a unified CPU-GPU snapshot.
Restoring the state of CUDA applications has the following steps:
\begin{enumerate}[label=\itshape(\roman*\upshape),nosep]
    \item \textit{Restore} resources such as device memory back to the GPU, memory mappings to their original addresses, and reconstruct CUDA objects (e.g., streams and contexts).
    \item \textit{Unlock} driver APIs, allowing the CUDA application to resume execution on the GPU.
\end{enumerate}
In addition, a boolean flag is set in the inventory image of the snapshot indicating whether it contains GPU state, allowing for compatibility checks and optimizations during restore.

\begin{figure*}[t]
    \centering
    \begin{subfigure}[]{0.45\textwidth}
        \centering
        \includegraphics[width=\textwidth]{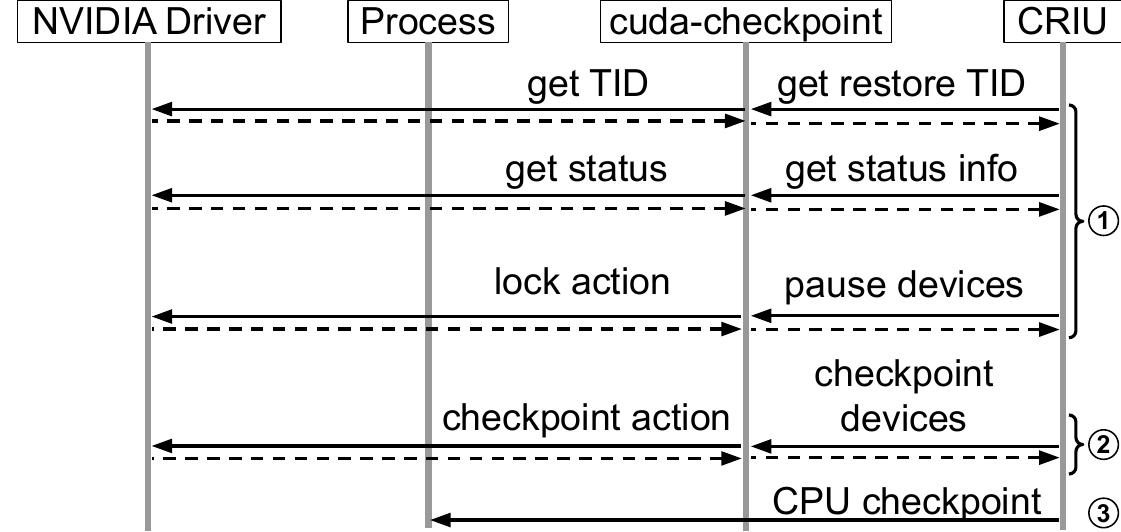}
        \caption{Sequence of interactions between CRIU, cuda-checkpoint, and NVIDIA driver.}
        \label{fig:cuda-checkpoint-flow}
    \end{subfigure}
    \hspace{0.95em}
    \rulesep
    \hspace{0.05em}
    \begin{subfigure}[]{0.45\textwidth}
        \centering
        \includegraphics[width=.85\textwidth]{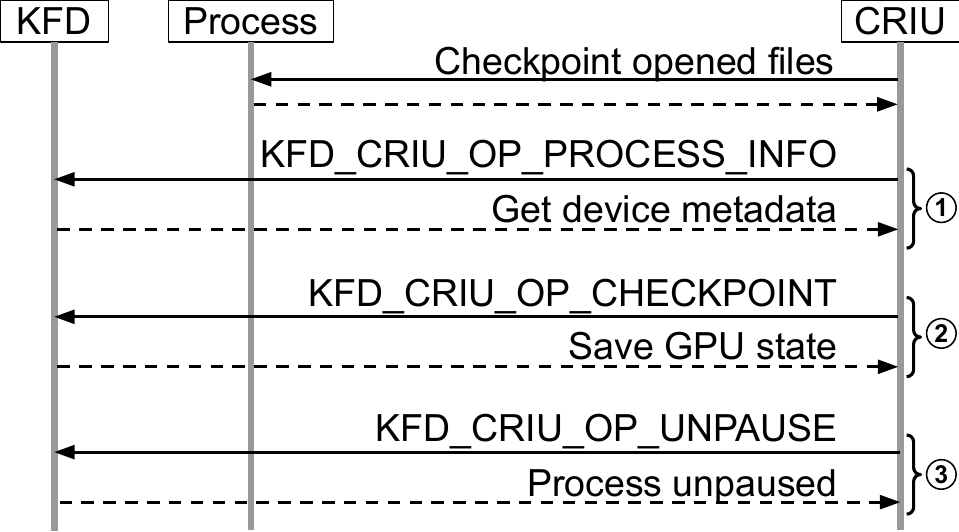}
        \caption{Sequence of interactions between CRIU and KFD.}
        \label{fig:amdgpu-checkpoint-flow}
    \end{subfigure}
    \vspace{-0.5em}
    \caption{Sequence diagrams of CRIU interactions with NVIDIA and AMD drivers.}
    \label{fig:plugins-checkpoint-workflow}
    \vspace{-1em}
\end{figure*}
\subsubsection{AMD GPU Plugin}\label{sec:amd-gpu-plugin}
The AMD GPU plugin enables transparent checkpoint/restore using input/output control (\texttt{ioctl}) operations with the Kernel Fusion Driver (KFD). These operations are used to pause and resume the execution of GPU processes, as well as to capture and restore their state, which consist of memory buffer objects (BOs), queues, events, and topology.

GPU-accessible BOs are kernel-managed device (VRAM) and system (graphics translation table) memory, user-managed memory (userptr), and special apertures for signaling (doorbell) and control registers (MMIO). The saved BO properties include buffer type, handle, size, virtual address, device file offset for CPU mapping, and memory contents.

GPU work is typically submitted through user-mode queues with associated user- and kernel-managed memory buffers.
Checkpointing requires preempting and saving the state of all queues belonging to the process.
This includes queue type (compute or DMA), kernel-managed control stack, memory queue descriptor, read/write pointers, doorbell offset, and architected queueing language (AQL) pointer.
The state stored in user-managed BOs includes ring buffer (commands), AQL queue, completion tracking (end-of-processing) buffer, and context save area (preempted shader state).
For checkpoint/restore of GPU-to-host signaling events, the allocated event IDs and their signaling state are saved and restored, while the event slot contents are included in the memory data.
During checkpointing, the plugin performs the following \texttt{ioctl} operations:
\begin{enumerate}[label=\itshape(\roman*\upshape), nosep]
    \item \texttt{PROCESS\_INFO} -- collecting metadata about the process, pausing its execution, and evicting all queues

    \item \texttt{CHECKPOINT} -- capturing the GPU state described above

    \item \texttt{UNPAUSE} -- restores the evicted queues
\end{enumerate}
For security reasons, KFD allows these \texttt{ioctl} calls to be performed only by the same process that opened the \texttt{/dev/kfd} file descriptor, and requires \texttt{CAP\_CHECKPOINT\_RESTORE} or \texttt{CAP\_SYS\_ADMIN} capability.
The plugin performs the following operations during restore:
\begin{enumerate}[label=\itshape(\roman*\upshape),nosep]
    \item \texttt{RESTORE}: reinstates the checkpointed state of processes
    \item \texttt{RESUME}: resumes execution of processes on the GPU
\end{enumerate}

Checkpointed applications can only be restored on systems with compatible GPU topology with the same number, type, memory size, VRAM accessibility by the host, and connectivity between GPUs.
When restoring on a different system or with different subset of GPUs on the same system, the unique GPU identifiers (GPUIDs) might be different during restore. These identifiers are based on properties like the instruction set and compute units. To address this, the plugin performs a translation of the GPUIDs used by the restored processes that applies to all KFD ioctl calls.

\subsubsection{Plugin Hooks}\label{sec:plugin-hooks}
CRIU provides a set of hooks for checkpointing external resources such as UNIX sockets, file descriptors, mountpoints, and network devices. These hooks serve as an API that can be used with plugins to extend the existing functionality.

\stitle{AMD GPU Plugin Hooks.} CRIU provides two hooks for handling checkpoint and restore operations with device files: \texttt{DUMP\_EXT\_FILE} and \texttt{RESTORE\_EXT\_FILE}. When checkpointing ROCm applications, these hooks are invoked for the \texttt{/dev/kfd} and \texttt{/dev/dri/renderD*} device nodes. The obtained KFD file descriptor is used by the plugin to perform \texttt{ioctl} calls to manage memory, queues, and signals, while per-GPU device render node files are utilized to handle CPU mapping of VRAM and GTT BOs. Two additional plugin hooks have been introduced to enable checkpoint/restore of AMD GPU device virtual memory areas (VMA): \texttt{HANDLE\_DEVICE\_VMA} and \texttt{UPDATE\_VMA\_MAP}. These hooks allow the plugin to translate device file names and mmap offsets to newly allocated ones during restore. In particular, these offsets identify BOs within a render node device file, and the translation mechanism allows a process to be restored on a different GPU. A \texttt{RESUME\_DEVICES\_LATE} hook has been introduced to finalize the restore of userptr mappings and resume execution on the GPU for each restored process, after CRIU's restorer PIE code has restored all VMAs.

\stitle{CUDA Plugin Hooks.} Similarly, two additional plugin hooks have been introduced to invoke \textit{lock} and \textit{checkpoint} actions with the \texttt{cuda-checkpoint} utility for processes running on NVIDIA GPUs: \texttt{PAUSE\_DEVICES} and \texttt{CHECKPOINT\_DEVICES}. The \textit{pause} hook is called immediately before the target CPU processes are frozen. This hook is used by the CUDA plugin to place the corresponding GPU tasks in a \textit{locked} state, halting any pending work and preparing them to be checkpointed. Following this, the \textit{checkpoint} hook is called after all CPU and GPU processes are frozen/locked state to checkpoint their GPU state into host memory. The CUDA plugin also utilizes the \texttt{RESUME\_DEVICES\_LATE} hook to \textit{restore} the state of processes from host memory to the GPU and perform the \textit{unlock} action to resume their execution.

\subsection{Checkpoint/Restore Workflow}
The sequence of operations described above for AMD GPU and CUDA plugins is illustrated in \Cref{fig:plugins-checkpoint-workflow}. Each plugin uses a different method for checkpointing the GPU state of applications. For CUDA applications, \circled{1} performs a \textit{lock} action that halts the execution of device API calls. Similarly, the AMD GPU plugin invokes a KFD ioctl call to collect metadata, pause execution, and the evict queues for the target ROCm application. \circled{2} checkpoints the GPU state to host memory for CUDA applications. In contrast, at this stage the AMD GPU plugin saves the GPU state into a set of checkpoint files. \circled{3} continues with traditional checkpoint operations for CUDA applications as the GPU state is included in host memory. The AMD GPU plugin at this stage invokes a KFD ioctl call to resume the state of queues. The restore functionality has analogous sequence operations as described in \Cref{sec:gpu-plugins}.
\section{Checkpointing GPU-Accelerated Containers}\label{sec:implementation}%
Integrating \sys with container runtimes like Podman, CRI-O and containerd involves leveraging the Container Device Interface (CDI)~\cite{cdi} and the NVIDIA Container Toolkit~\cite{container-toolkit} to enable running containers with GPU hardware. These container runtimes already support the checkpoint/restore of containerized applications using the CRIU tool. However, to enable checkpoint/restore of GPU-accelerated containers, further modifications are necessary.

\subsection{Checkpoint/Restore of GPU Containers}
Accessing GPU resources in containers requires a set of utilities that automatically generate configuration to expose within the container the appropriate device files and libraries. This runtime configuration typically includes details such as the number of GPUs, their unique identifiers, and \textit{driver capabilities} (e.g., compute, graphics, video, utility, display). Runtime libraries such as libnvidia-container provide additional capabilities to container runtimes that enable configuring a container with GPU support by exposing device drivers to it. This functionality enters the container mount namespace referred by its root process identifier (PID) and performs a set of actions to make available GPU device files, libraries and utilities inside the container. The runtime library assumes that the container root filesystem has been created and the container is not yet started. As CRIU checkpoints the mount namespace of the container, it also needs to handle all external mounts created by the runtime library. This requires the runtime library to update the container configuration with all new mounts, enabling the container runtime to specify the necessary CRIU options during checkpoint and restore.

\subsection{Multiprocess GPU Checkpoints}
Containers typically run multiple processes that are isolated from other containers and the host system with Linux control groups (cgroups) and namespaces. Creating a consistent checkpoint of the container process tree requires temporarily suspending the execution of all processes on both the CPU and the GPU. This mechanism is commonly implemented with the \textit{freezer} cgroup in combination with \textit{seize} and \textit{interrupt} functionality of the \texttt{ptrace} system call. In addition, the \textit{lock} and \textit{unlock} functionality of the \texttt{cuda\_checkpoint} tool is used to block/unblock all CUDA driver API calls of the running process. However, locking the driver API calls of all processes in a container is challenging because it requires these processes to be in a running state. Attempting to lock a frozen GPU application causes the \texttt{cuda\_checkpoint} tool to block until the process is resumed. To solve this problem, \sys avoids the use of the freezer cgroup when checkpointing CUDA applications. Instead, only the ptrace \textit{seize} and \textit{interrupt} mechanism is used to suspend the execution on the CPU. This functionality is implemented as part of the CUDA plugin. When the plugin is enabled, it configures CRIU to interrupt the execution of processes without the freezer cgroup.

\subsection{Filesystem Snapshots}
Container engines typically utilize a layered filesystem where container images are mounted as read-only layers to form the container root filesystem (rootfs), and a writable layer is mounted on top of these layers to enable creating files and modifying existing ones. This writable layer captures all changes made within a running container. Since the read-only layers are immutable, the checkpointing mechanism needs to capture only the changes stored in the writable layer. To create a consistent checkpoint, container engines typically utilize the \textit{cgroup freezer} to suspend all processes in a container. This allows saving their runtime state to persistent storage and the container engine to create a consistent copy of the rootfs writable layer. However, as mentioned above, this freeze mechanism prevents the \texttt{cuda-checkpoint} tool from locking the GPU threads. To address this problem, \sys identifies if the process tree is in a frozen state, and if it is, it resumes its execution and uses the ptrace seize with interrupt mechanism. It then uses the CPU-GPU checkpointing workflows (as illustrated in \Cref{fig:cuda-checkpoint-flow}), and leaves the container in a frozen state after the checkpoint has been created. The container runtime then captures the rootfs changes as described above.

\subsection{Checkpointing with NVML}
To enable checkpointing for CUDA applications using NVML, such as PyTorch, we extend the CUDA plugin to handle leftover device references that are not currently supported for checkpointing by the CUDA driver. In most cases, these device references are used during initialization to obtain information such as the available number of GPUs and their capabilities. These values are unlikely to change during runtime and are typically not accessed again. In addition, the \texttt{sys} restore operation requires the same GPU type and order to be used as during checkpointing. For example, if a checkpoint is created on a 4-GPU A100 system, it must be restored on another 4-GPU A100 system. Attempting to restore the checkpoint on an 8-GPU A100 system or a 4-GPU system with different types of GPUs will result in failure. Thus, we implement \texttt{DUMP\_EXT\_FILE}, \texttt{RESTORE\_EXT\_FILE}, \texttt{HANDLE\_DEVICE\_VMA}, and \texttt{UPDATE\_VMA\_MAP} plugin hooks to ensure proper handling of device references and memory mappings during the checkpoint and restore operations. A future release of the CUDA Display Driver will provide additional checkpointing capabilities simplifying this implementation.

%
\begin{table*}[t]
\centering
\setlength{\tabcolsep}{3pt} 
\renewcommand{\arraystretch}{1} 
\begin{tabular}{@{}llr|r|r@{}} 
\toprule
\textbf{GPUs} & \textbf{CPU} & \textbf{CPU cores} & \textbf{Memory} & \textbf{Storage} \\ 
\midrule
H100 (PCIe 5.0 80GB HBM3) & Intel Xeon Platinum 8458P (2.70GHz) & 16 & 256~GB & 2~TB (HDD) \\ 
1x~/~2x~/~4x~A100 (SXM4 80GB) & Intel Xeon Gold 6342 (2.80GHz) & 12 / 24 / 48 & 96/180/368~GB & 2~TB (HDD) \\ 
1x~/~2x~/~4x~V100 (SXM2 32GB) & Intel Xeon Gold 6130 (2.10GHz) & 8 / 16 / 32 & 32 / 64 / 120~GB & 1~TB (HDD) \\ 
1x~/~2x~/~4x~RTX A6000 & Intel Xeon Gold 5315Y (3.20GHz) & 8/16/32 & 44 / 90 / 180~GB & 1~TB (HDD) \\ 
MI210 (PCIe 4.0 64GB HBM2e) & Intel Xeon Gold 6342 (2.80GHz) & 48 & 512~GB & 1.5~TB (NVMe) \\ 
\bottomrule
\end{tabular}
\vspace{-.5em}
\caption{Specifications of the 10 servers equipped with NVIDIA GPUs and one with an AMD GPU used in the evaluation.}
\label{tab:server-configs}
\vspace{-1em}
\end{table*}
%
\begin{table}[t]
\centering
\setlength{\tabcolsep}{3pt} 
\renewcommand{\arraystretch}{1.1} 
\begin{tabular}{@{}lcccc@{}}
\toprule
\multicolumn{1}{c}{\multirow{2}{*}{\textbf{Time}}}
& \multicolumn{2}{c}{\textbf{Llama 3.1 (8B)}} & \multicolumn{2}{c}{\textbf{GPT2-XL (1.5B)}} \\
\cmidrule(lr){2-3} \cmidrule(lr){4-5}
& \textbf{H100} & \textbf{A100} & \textbf{H100} & \textbf{A100} \\
\midrule
CPU Freezing (\textit{s}) & 26.85 & 24.31 & 30.30 & 25.94 \\
CPU Frozen (\textit{s}) & 50.54 & 122.11 & 58.50 & 104.85 \\
CPU Mem. dump (\textit{s}) & 48.88 & 119.79 & 56.75 & 102.27 \\
CPU Mem. write (\textit{s}) & 47.24 & 117.60 & 54.80 & 99.81 \\ \midrule
\sys Checkpoint (\textit{s}) & 77.40 & 146.43 & 88.81 & 130.81 \\
\sys Restore  (\textit{s}) & 38.83 & 98.91 & 43.43 & 145.14 \\
\midrule
GPU memory & 54 GB & 54 GB & 58 GB & 58 GB \\
\bottomrule
\end{tabular}
\vspace{-.5em}
\caption{\sys checkpoint and restore times for Llama 3.1 and GPT-2 model training on a single H100 80GB and A100 80GB GPU. While H100 architectural improvements and faster memory bandwidth enable faster checkpoint/restore, the additional CPU cores and host memory likely also contribute to the overall speedup. A direct comparison with A100 is not possible due to these differences.}
\label{tab:combined-checkpoint-restore-times}
\end{table}
\subsection{Multi-GPU Checkpoints}
\sys supports applications running on a single-node with multiple GPUs. To ensure consistency of concurrent operations, transparent checkpoint/restore for multi-GPU applications requires identifying all associated tasks and synchronizing the state across all GPUs. Multi-GPU applications often perform computations with methods implementing data parallelism at the module level. For example, in the case of training jobs a module is parallelized by replicating it across devices, splitting the input batch dimension among them, and aggregating gradients during back-propagation. The \texttt{cuda-checkpoint} tool provides \textit{lock} and \textit{unlock} actions that enable synchronization across multiple CUDA tasks. These actions are utilized by the CUDA plugin to pause and resume task execution on GPU devices. The CUDA driver handles the internal GPU state during checkpoint and restore operations to/from host memory to ensure complete and consistent snapshot state.
Distributed checkpointing for multi-node workloads requires additional synchronization mechanisms with both CPU and GPU state across different nodes. Applications and frameworks typically use libraries such as NVIDIA Collective Communications Library (NCCL) to implement this functionality. However, at the time of writing, the \texttt{cuda-checkpoint} tool does not support checkpoint/restore operations with NCCL. This functionality is expected to become available in a future release of the CUDA driver.

\begin{table*}[t]
\centering
\begin{tabular}{@{}l|ccc|ccc|ccc@{}}
\toprule
\multicolumn{1}{c} {\textbf{Time} (\textit{s})}&
  \multicolumn{1}{l}{\textbf{4xA100}} &
  \multicolumn{1}{l}{\textbf{2xA100}} &
  \multicolumn{1}{l}{\textbf{A100}} &
  \multicolumn{1}{l}{\textbf{4xA6000}} &
  \multicolumn{1}{l}{\textbf{2xA6000}} &
  \multicolumn{1}{l}{\textbf{A6000}} &
  \multicolumn{1}{l}{\textbf{4xV100}} &
  \multicolumn{1}{l}{\textbf{2xV100}} &
  \multicolumn{1}{l}{\textbf{V100}} \\ \midrule
CPU Freezing (\textit{s}) & 21.49 & 10.32 & 4.96  & 15.23 & 9.11  & 3.35  & 29.41  & 14.50 & 6.90  \\
CPU Frozen (\textit{s}) & 33.58 & 16.15 & 7.79  & 43.69 & 29.48 & 11.24 & 74.96  & 38.56 & 19.23 \\
CPU Mem. dump (\textit{s}) & 31.30 & 15.02 & 7.28  & 42.1  & 28.59 & 10.88 & 70.30  & 36.17 & 18.10 \\
CPU Mem. write (\textit{s}) & 28.62 & 13.99 & 6.80  & 40.4  & 27.70 & 10.49 & 66.30  & 34.40 & 17.26 \\ \midrule
\sys Checkpoint (\textit{s}) & 55.09 & 26.49 & 12.78 & 58.93 & 38.61 & 14.60 & 104.40 & 53.08 & 26.18 \\
\sys Restore (\textit{s}) & 35.13 & 17.22 & 8.32  & 24.1  & 13.83 & 5.50  & 43.14  & 21.69 & 10.61 \\ \midrule
Checkpoint size (GB) & 41.01 & 20.46 & 9.94 & 39.98 & 19.97 & 9.75 & 40.03  & 19.97 & 9.81 \\ \bottomrule
\end{tabular}\par
\vspace{-0.5em}
\captionsetup{justification=centering}
\caption{Checkpoint and restore performance (in seconds) when scaling training of GPT-2 Small (124M) to multiple GPUs.\\Times are not comparable across GPU families.}
\label{tab:multi-gpu-checkpointing}
\end{table*}

\begin{figure}[t]
    \centering
    \includegraphics[width=.9\columnwidth]{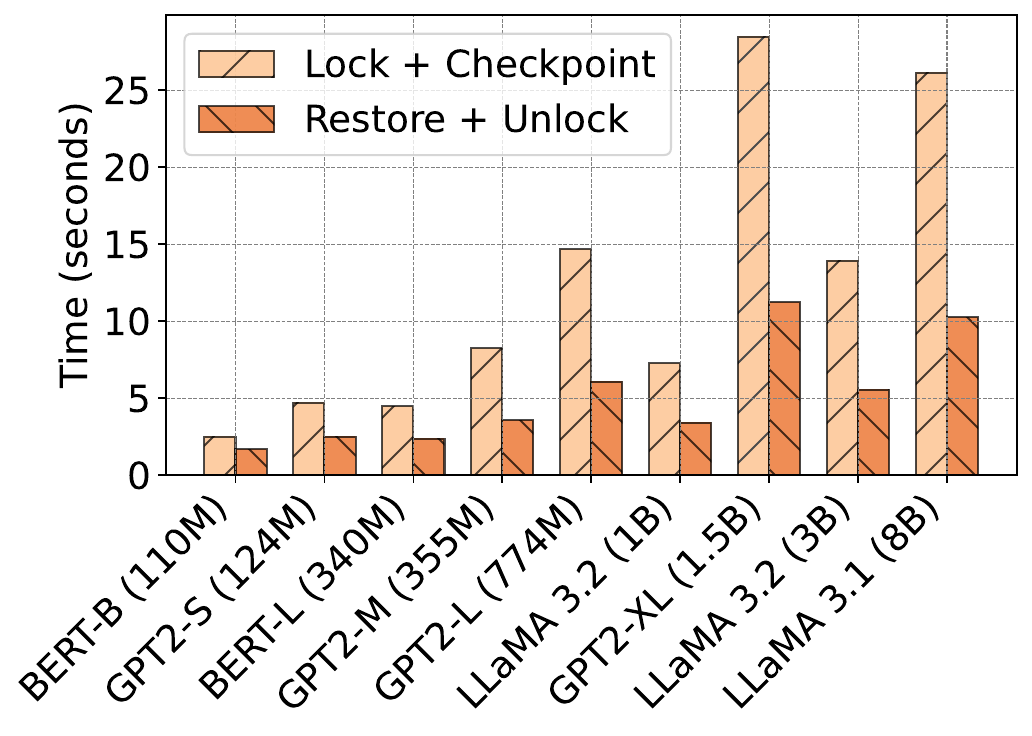}
    \vspace{-.5em}
    \caption{In-memory GPU checkpoint/restore with H100. Similar results are observed with A100.}
    \label{fig:in-memory-checkpoint-restore}
\end{figure}
\begin{figure*}[t]
  \centering
  \begin{subfigure}[b]{\columnwidth}
    \includegraphics[width=.9\textwidth]{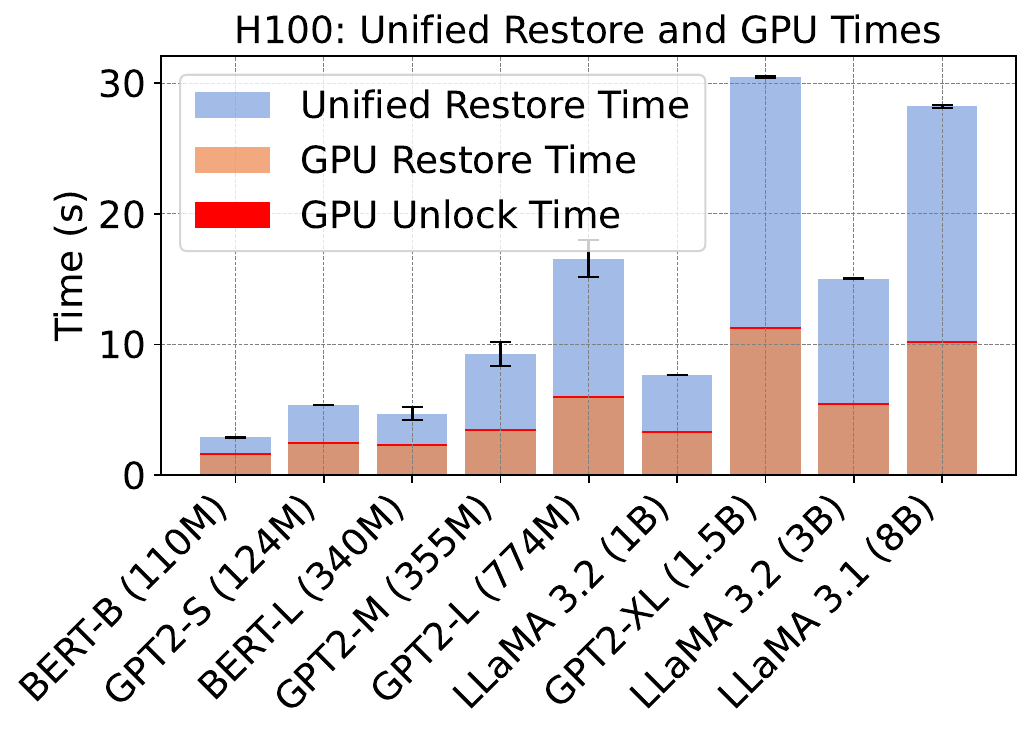}
    \label{fig:h100-unified-restore-times}
  \end{subfigure}
  \hfill
  \begin{subfigure}[b]{\columnwidth}
    \includegraphics[width=.9\textwidth]{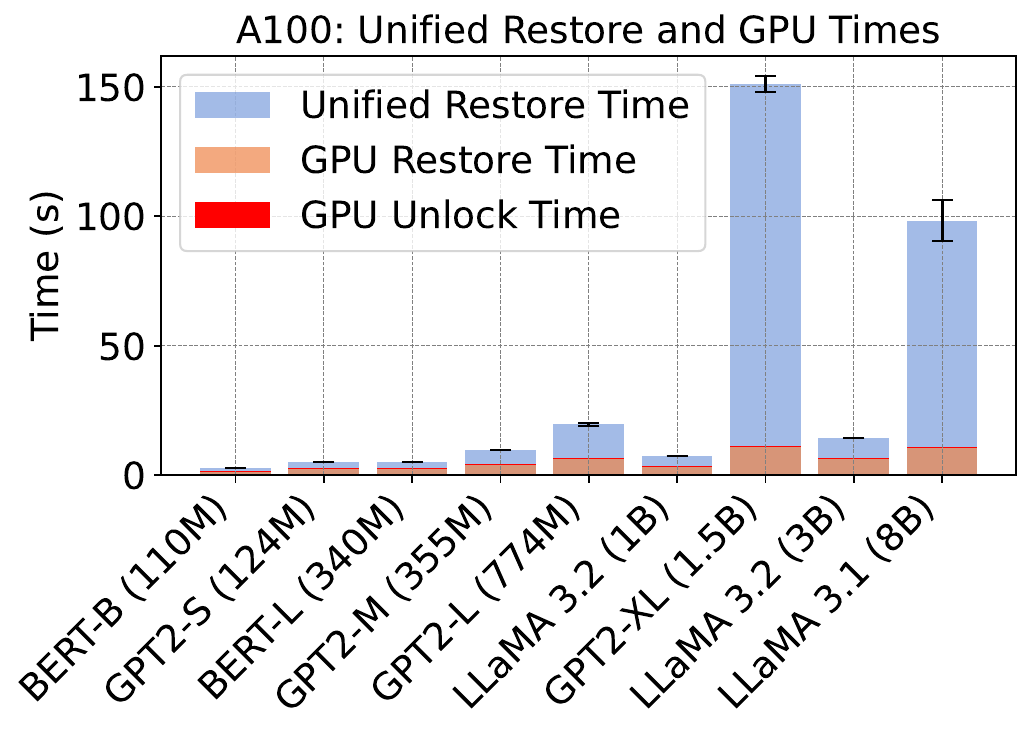}
    \label{fig:a100-unified-restore-times}
  \end{subfigure}
  \vspace{-1em}
  \caption{Time to restore for model training from a checkpoint with \sys for H100 and A100 GPUs.}
  \label{fig:unified-restore-times}
\end{figure*}

\section{Evaluation} \label{sec:evaluation}%
Our evaluation seeks to answer the following questions:
\begin{itemize}[leftmargin=*,leftmargin=15pt,itemindent=0pt]
    \item How does \sys perform when checkpointing and restoring large language models? (\textsection{\ref{sec:eval:diff-models}})

    \item What are the scalability implications of using checkpointing and restoring with multiple GPU devices? (\textsection{\ref{sec:eval:scalability}})

    \item What are the dominant factors affecting the latency of checkpointing and restore operations? (\textsection{\ref{sec:eval:overhead}})

    \item Can \sys support checkpoint and restore with both CUDA and ROCm applications? (\textsection{\ref{sec:eval:rocm}})
\end{itemize}

\subsection{Experimental Methodology}%
\stitle{Evaluation Setup.}
We evaluate \sys on 10 servers with specifications described in \Cref{tab:server-configs}, running Ubuntu 22.04 with kernel version 6.2.0 (A100 and H100), 5.15.0 (V100 and A6000), and NVIDIA driver 565.57.01, CUDA 12.7, and CentOS Stream 9 with kernel 5.14 with ROCm 5.6 (MI210).

\stitle{Performance Measurements.} We measure the performance of checkpoint and restore operations using detailed statistics generated by CRIU~\cite{criu-statistics} about the time spent in different stages, and with external tools like \texttt{perf stat} to gather more detailed performance data. We run each experiment 10 times and calculate the mean and standard deviation of each value in the collected data. To analyze the overhead of checkpointing with \sys, we measure the following performance metrics for models of different sizes:

\begin{itemize}[leftmargin=*,leftmargin=10pt,itemindent=0pt]
    \item \textbf{Checkpoint time:} The total time to create a snapshot of the running GPU application.
    \item \textbf{Freezing time:} The time to suspend the application using \texttt{ptrace} seize and interrupt.
    \item \textbf{Frozen time:} The time during checkpointing when the application is not running.
    \item \textbf{Memory dump time:} The time to collect the CPU memory pages of running processes. This does not include the time to write this memory to storage.
    \item \textbf{Memory write time:} The time to save the memory state to persistent storage.
    \item \textbf{Restore time:} The time to restore both CPU and GPU state from storage, and to resume the application.
\end{itemize}

\stitle{Workloads and micro-benchmarks.} We evaluate the proposed checkpoint/restore mechanisms for multiple models of different sizes (listed below) and a set of ROCm micro-benchmarks~\cite{amd2024rocm} representing common HPC workloads for AMD GPU~(\textsection{\ref{sec:eval:rocm}}). For NVIDIA GPUs we use the following models:

\begin{itemize}[leftmargin=*,leftmargin=10pt,itemindent=0pt]
    \item \textbf{LLaMA} \textbf{3.2} (1B, 3B) and \textbf{3.1} (8B)

    \item \textbf{GPT-2} with 124M, 355M, 774M, 1.5B parameters

    \item \textbf{BERT} Base (110M) and Large (340M) models
\end{itemize}
\subsection{\sys Performance with Deep Learning Models}
\label{sec:eval:diff-models}
We evaluate the performance of GPU checkpoint and restore operations with multiple model training workloads of different sizes. The results in~\Cref{fig:in-memory-checkpoint-restore} show that the time required to checkpoint the GPU state into host memory increases significantly for models with large number of parameters.
For instance, checkpointing and locking operations for the GPT-2 Small model (124M parameters) take an average of 4.9 seconds and 240 ms, respectively. In comparison, for a larger model such as GPT-2 XL (1.5B parameters), these operations require an average of 28 seconds and 500 ms, respectively.
The time to restore the GPU state from host memory increases gradually, with 2.5 seconds for GPT-2 Small and 11 seconds for GPT-2 XL, while the unlock time remains consistent for both models at approximately 160 ms.
We observe similar results with both A100 and H100 GPUs, highlighting the crucial impact memory bandwidth has on the performance of GPU checkpointing operations.
Several techniques have been proposed to address this problem, such as data compression and on-demand parallelism~\cite{yang2024on-demand}. Incorporating such techniques could further improve the efficiency of checkpoint/restore operations, especially for large-scale models.

~\Cref{fig:unified-restore-times} shows the unified restore time (the time to restore the combined CPU-GPU state) for models with different sizes with both H100 and A100 GPUs. The time required to restore the GPU state makes up a significant portion of the total restore time for small models, but becomes a relatively lesser portion for larger models. These results demonstrate that the restore time is also affected by the available bandwidth for CPU-GPU memory transfers, as well as the speed at which checkpoint data is loaded from disk into host memory.
The performance results and checkpoint sizes shown in \Cref{tab:combined-checkpoint-restore-times} suggest that the differences in performance between the experiments with H100 and A100 GPUs can be attributed not only to advancements in GPU hardware architecture but also to the critical role of the available CPU resources.

\begin{table}[t]
\centering
\renewcommand{\arraystretch}{.9} 
\begin{tabular}{@{}c|rrr@{}}
\toprule
\textbf{Model} & \textbf{Total (GB)} & \textbf{GPU (\%)} & \textbf{CPU (\%)} \\ \midrule
BERT-B~~(110M)  & 5.22  & 82.38\%  & 17.62\% \\
GPT2-S~~(124M)  & 10.32 & 89.15\%  & 10.85\% \\
BERT-L~~(340M)  & 9.90  & 90.91\%  & 9.09\%   \\
GPT2-M~(355M)  & 19.57 & 91.31\% & 8.69\%   \\
GPT2-L~~(774M)  & 35.39 & 90.99\% & 9.01\%  \\
LLaMA 3.2~~~(1B) & 14.81 & 92.37\% & 7.63\%  \\
LLaMA 3.2~~~(3B) & 29.54 & 95.70\%  & 4.30\%   \\
LLaMA 3.1~~~(8B) & 55.89 & 97.35\% & 2.65\%  \\
GPT2-XL~(1.5B) & 60.12 & 96.02\% & 3.98\%  \\ \bottomrule
\end{tabular}
\vspace{-.5em}
\caption{The total unified checkpoint size (in GB) and the corresponding proportions of GPU memory and CPU state, respectively, for various models running on H100 GPU. We observe similar results with A100 GPU.}
\label{tab:cpu-to-gpu-state-comparison}
\vspace{-1.5em}
\end{table}

\Cref{tab:cpu-to-gpu-state-comparison}  shows the total \sys checkpoint sizes for various models on both A100 and H100 GPUs, along with a breakdown of GPU memory and CPU state proportions. A key insight is the dominance of GPU memory in overall checkpoint size, consistently exceeding 80\% and often surpassing 90\% for larger models like LLaMA 3.1 (8B) and GPT2-XL (1.5B). These results further emphasize the importance of efficient CPU-GPU memory transfers in optimizing the performance of checkpointing and restore operations.

\subsection{Multi-GPU Checkpointing Performance}
\label{sec:eval:scalability}
We evaluate the scalability of \sys by running an experiment designed to train a large language model (GPT-2) across 1x, 2x, and 4x GPUs of V100, A6000, and A100 types, using data parallelism to distribute the workload.
\Cref{tab:multi-gpu-checkpointing} shows that the checkpoint size increases with the number of GPUs, as each GPU stores its own copy of the model parameters.
For instance, the checkpoint size for 1, 2, and 4 A100 GPUs is $\approx10$, $\approx20$, and $\approx40$ GB, respectively.
This increase in checkpoint size reflects the increasing amount of intermediate model state that is saved as the number of GPUs increases.
The freezing, frozen, memory dump and memory write times also increase with the number of GPUs, likely because more time is spend on handling the larger checkpoint data with additional GPUs.
For example, as the number of A100 GPUs increases, creating a unified CPU-GPU snapshot requires scanning more memory pages.
A single GPU requires $\approx7$ million pages scanned, two GPUs require $\approx15$ million, and four GPUs require $\approx28$ million. Our experiments demonstrate that \sys efficiently scales as the number of GPUs and data-parallel replicas increase. We observe that checkpointing and restore times scale near linearly as we increase the number of GPUs from 1 to 4 across different GPU types.

\subsection{Checkpoint and Restore Latency}%
\label{sec:eval:overhead}%
We analyze the overhead of checkpoint and restore operations of \sys by measuring the latency during these processes for LLaMA 3.1 and GPT2-XL model training workloads on H100 and A100 GPUs. The results in ~\Cref{tab:combined-checkpoint-restore-times} highlight the performance differences between the H100 and A100 GPUs, as well as the impact of additional CPU and memory resources on reducing overhead during the checkpoint and restore operations with \sys. The primary factors affecting \sys's checkpoint and restore performance are:

\begin{itemize}
    \item \textbf{GPU count}: Increasing the number of GPUs leads to increased checkpoint size and latency;
    \item \textbf{CPU-GPU bandwidth}: The speed of data transfers between the CPU and GPUs directly affects checkpointing and restoring speed;
    \item \textbf{GPU memory usage}: Larger models with more parameters have higher checkpoint and restore latencies.
\end{itemize}

\begin{figure}[t]
  \centering
  \includegraphics[width=\columnwidth]{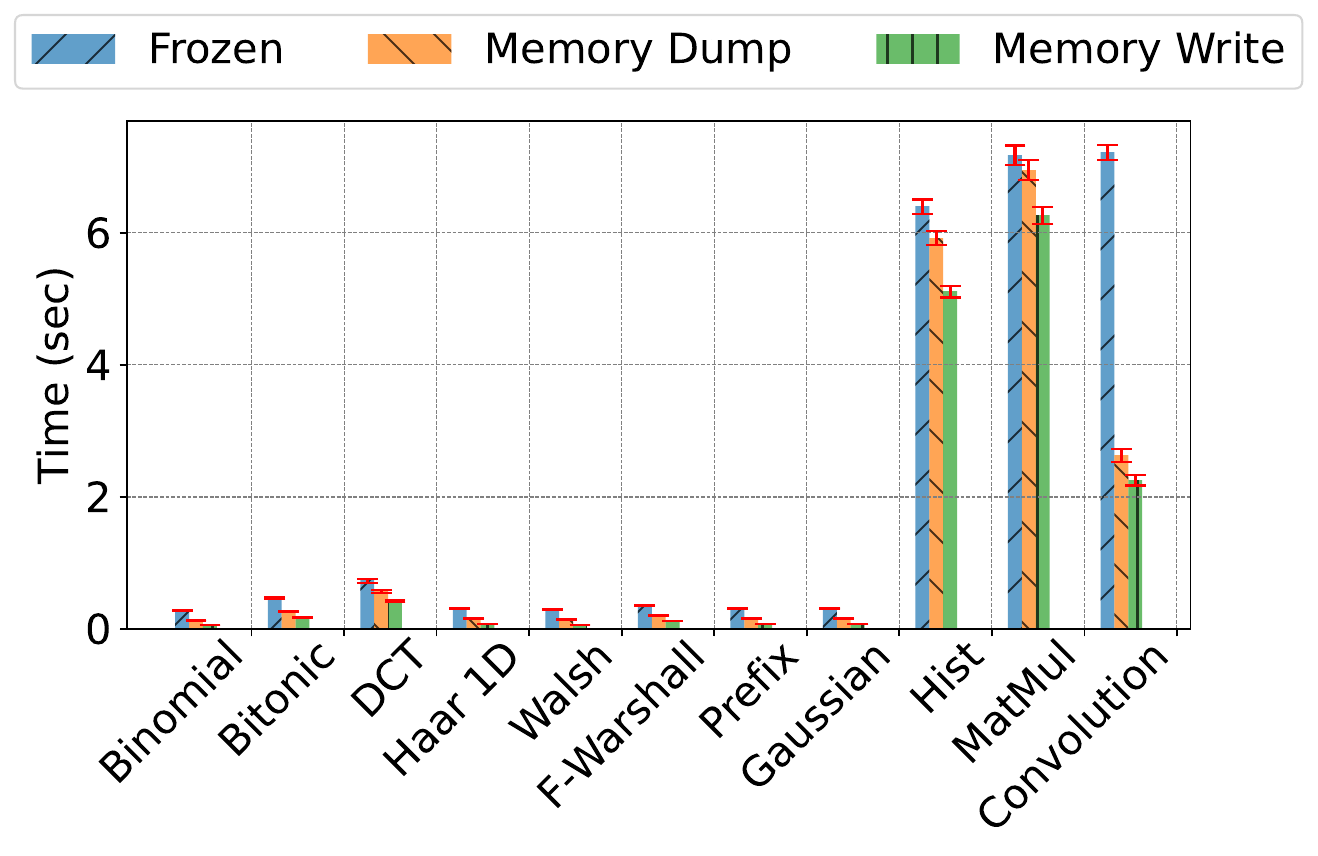}
  \vspace{-2.5em}
  \caption{Breakdown of the \sys checkpointing time for HPC benchmarks running on AMD MI210 GPU.}
  \label{fig:amd-gpu-checkpointing}
  \vspace{-.5em}
\end{figure}
\begin{table}[t]
\centering
\begin{tabular}{lr}
\toprule
\textbf{Benchmark} & \textbf{Checkpoint Size}
\\ \midrule
Binomial Option Pricing       & 305~MB \\
Bitonic Sort                  & 614~MB \\
Discrete Cosine Transform     & 1.2~GB \\
1D Haar Wavelet Decomposition & 333~MB \\
Fast Walsh Transform          & 307~MB \\
Floyd Warshall                & 484~MB \\
Prefix Sum                    & 306~MB \\
Recursive Gaussian            & 311~MB \\
Histogram                     & 16.64 GB \\
Matrix Multiplication         & 19.88 GB \\
Convolution                   & 13.83 GB \\
\bottomrule
\end{tabular}
\caption{\sys checkpoint sizes of ROCm benchmarks.}
\label{tab:rocm-benchmark-checkpoints}
\vspace{-1em}
\end{table}

\subsection{\sys Support for ROCm Devices}
\label{sec:eval:rocm}

In addition to support for CUDA, to demonstrate the checkpointing functionality of \sys for AMD GPUs, we evaluate its performance using a set of ROCm HPC micro-benchmarks. These benchmarks provide a set of workloads representative of typical HPC applications, allowing us to analyse \sys's ability to effectively checkpoint and restore GPU state across different computational patterns. \Cref{fig:amd-gpu-checkpointing} shows the frozen, memory dump, and memory write times during checkpointing for each benchmark. While most of the evaluated benchmarks have relatively small checkpoint size, typically ranging from under 500~MB to 1.2~GB, a few have significantly larger checkpoint sizes (Histogram, Matrix Multiplication, and Convolution), shown in \Cref{tab:rocm-benchmark-checkpoints}. This increase in checkpoint size directly correlates with longer freezing and memory dump times, as \sys must checkpoint larger amount of data. An interesting observation is the contrasting distribution of checkpoint data between host memory and GPU memory across these benchmarks. While more than half of the checkpoint size for the Convolution benchmark is attributed to AMD GPU state, Histogram and Matrix Multiplication have the majority of their state residing in host memory.

\section{Discussion}~\label{sec:discussion}%
\sys is a novel technique for fully transparent checkpointing of GPU applications implemented with CUDA and ROCm. Our evaluation results show that \sys can create unified CPU-GPU snapshots for large models in single- and multi-GPU setups without steady-state performance overhead. Our evaluation results further demonstrate that \sys scales linearly with the number of GPU devices for data-parallel workloads, enabling efficient checkpointing with large-scale applications.

\stitle{Deterministic Restore.}
In contrast to previous work, which requires validation of the replayed GPU device API calls~\cite{gupta2024just}, \sys relies on a locking mechanism that ensure the execution of tasks is suspended before creating a checkpoint. This guarantees consistent CPU-GPU snapshots and deterministic restore operations, during which both CPU and GPU state are restored before the application resumes execution.

\stitle{Real-World Deployments.}
The proposed GPU checkpointing mechanism has been successfully implemented in several production systems, including MemVerge~\cite{wu2025memory} and Modal~\cite{belotti2025memory}.

\section{Related Work}~\label{sec:related-work}%
Checkpoint/Restore has recently received significant attention in the context of AI workloads as a way to quickly recover from faults during model training (either in the GPU or in the CPU), and to better utilize hardware when multiple users content for the same hardware. The latter is becoming increasingly relevant with the introduction of LLM Agents in which GPU devices may need to wait for external agents to return before the LLM running on the GPU could proceed~\cite{nvidia2023introduction}. This section summarizes recent research efforts and how they have focused on improving different aspects of Checkpoint Restore, and how it compares to \sys.

\stitle{Checkpoint Types.} Distributed runtime engines such as Ray~\cite{moritz2018ray} and deep learning frameworks and libraries such as PyTorch~\cite{paszke2019pytorch}, TensorFlow~\cite{abadi2016tensorflow}, and MXNet~\cite{chen2015mxnet} allow users to specify checkpoint and restore logic including how and when to checkpoint. Such techniques are specific to each runtime engine and require developer involvement, thus being considered \textit{application-aware checkpoints}.

Many other works proposed and improved designs for transparent checkpoints. Singularity~\cite{shukla2022singularity} proposed a device proxy server design that relies on API interception to log all runtime and drive API calls. This log would be used to reconstruct the GPU state upon a restore operation. Such technique has been used widely as a mechanism for semi-transparent checkpoint restore support~\cite{chaudhary2020balancing,gupta2024just}. However, a design based on a device proxy server leads to a number of challenges (as discussed in \autoref{sec:background}), resulting in limited support for large and complex training and inference workloads. \sys, on the other hand, introduces a new design for checkpoint and restore by offering a fully-transparent and unified checkpoint mechanism to save the state of the application running on the CPU (including the engine/framework/library running the user application), and its corresponding state on the GPU. \sys does not rely on API interception and supports different GPU families with both CUDA and ROCm applications.

\stitle{Deciding when to Checkpoint.} Transparent GPU checkpointing is often used to enable error recovery for training jobs by periodically saving the model parameters and optimizer state to persistent storage. State-of-the-art error recovery solutions today implement \textit{periodic} or \textit{just-in-time} checkpointing support where training jobs can be resumed from any prior checkpoint. When GPU workloads experience an error, they either hang or crash, and cause the training jobs to be unexpectedly terminated. These errors are often detected with a series of diagnostic tests and problematic nodes are cordoned off, allowing the training job to resume on a different node.

Recovery mechanisms used with periodic checkpoints are often optimized to balance low running time overhead and high checkpoint frequency~\cite{mohan2021checkfreq}. These optimizations aim to minimize both the runtime overhead and the amount of work the job has to redo on recovery (recovery cost).
In contrast, just-in-time checkpointing solves this problem by creating a checkpoint only after a failure has occurred by leveraging the fact that the GPU state is only updated during a short interval and multiple replicas are holding the same state. As a result, this approach is able to recover from errors by recomputing at most one mini-batch~\cite{gupta2024just}.

\sys can be integrated with both periodic and just-in-time checkpointing policies to i) accelerate the checkpoint and restore operations, and ii) create a unified and consistent snapshot of the CPU and GPU state. Furthermore, \sys does not rely on specific workload characteristics such as relying on replicas having an exact copy of the model, thus being a general approach for GPU state checkpointing.

\stitle{Checkpoint/Restore Optimizations.} There have been many other works that optimize checkpoint/restore operations for GPU-accelerated workloads.
CheckFreq~\cite{mohan2021checkfreq} reduces the overhead of periodic checkpointing by overlapping communication with snapshotting of the model state in the GPU. 
Gemini~\cite{team2024gemini} checkpoints the GPU state to local and remote host memory, and interleaves checkpointing traffic with training traffic to reduce the overheads of checkpointing. 
Nebula~\cite{microsoft2024boost} copies model state asynchronously reducing the time during which the training job is paused for checkpoint. 
DeepFreeze~\cite{nicolae2020deepfreeze} proposes a fine-grain asynchronous checkpointing of deep learning models and shards the checkpointing effort across multiple workers. However, it only considers CPU clusters and does not take into account the cost of snapshotting the model state in memory when trained with state-of-the-art GPUs.
Check-N-Run~\cite{eisenman2022check} proposes incremental/differential checkpointing to speedup recommendation model training and uses quantization techniques to reduce the snapshot size.
ZeRo~\cite{rajbhandari2020zero} shards model parameters and optimizer state across data-parallel GPUs, parallelizing the checkpoint effort.

We consider our work to be complementary to these previous contributions. Optimizations based on sharding, compression, asynchronous, or incremental checkpointing could be incorporated into \sys. In summary, \sys presents a novel checkpoint/restore technique on which many of previously proposed optimizations could be integrated.

Another recent line of related work explores running GPU-accelerated inference workloads on serverless platforms, focusing on optimizing resource efficiency, scheduling, and caching policies~\cite{10.5555/3433701.3433792,280768,ishakian2018serving,280704,10.1145/3620678.3624664,infaas,10.1145/3503222.3507709,234998,yang2024on-demand}. Yang et al.~\cite{yang2024on-demand} also leverages CRIU to accelerate the restore time for GPU workloads, but, unlike \sys, does not explore unified container snapshots. Besides, the paper~\cite{yang2024on-demand} focuses on AMD-only devices and does not explore the performance of checkpoint and restore operations on large ML models.
Nevertheless, most of these systems focus on reducing model load time rather than on a general approach for checkpoint and restore GPU-accelerated applications. Many additional optimizations could be combined with \sys to further facilitate model loading. These optimizations are outside the scope of this paper and will be addressed in future work.

\stitle{Framework-level checkpointing} is supported by many popular machine learning libraries, such as PyTorch~\cite{paszke2019pytorch} and TensorFlow~\cite{abadi2016tensorflow}. This approach requires the application, framework or library to correctly capture and restore the runtime state of applications. However, implementing and maintaining this checkpointing mechanism introduces additional development and testing burdens. In addition, it has been shown to be both error-prone and inefficient, often leading to checkpoint file loss or corruption due to job interruptions, as well as significant recovery times~\cite{mohan2021checkfreq}. Thus, in this work, we focus on developing a reliable and efficient system-level checkpointing that is fully transparent.

\section{Conclusion}
\sys is, to the best of our knowledge, the first mechanism offering unified and fully-transparent CPU-GPU snapshots. With \sys, we further build transparent GPU-accelerated container snapshots, allowing GPU-accelerated and containerized applications to easily benefit from checkpoint/restore. We benchmark our prototype implementation on a wide range of ML models and hardware configurations and show that it is capable of efficiently checkpointing and restoring CPU and GPU state even on large language models.

\section*{Acknowledgments}
The authors gratefully acknowledge Felix Kuehling, Rajneesh Bardwaj, David Yat Sin, David Francis, and Ramesh Errabolu for their invaluable help and support with implementation of AMD GPU plugin. We acknowledge the use of the University of Oxford Advanced Research Computing (ARC) facility~\cite{richards2015university} in conducting parts of the evaluation experiments. This work is partly supported by the EPSRC Doctoral Training Partnership (DTP) [grant number EP/T517811/1].

\balance
\bibliographystyle{plain}
\bibliography{main}

\begin{thebibliography}{10}

\bibitem{abadi2016tensorflow}
Mart{\'\i}n Abadi, Paul Barham, Jianmin Chen, Zhifeng Chen, Andy Davis, Jeffrey Dean, Matthieu Devin, Sanjay Ghemawat, Geoffrey Irving, Michael Isard, Manjunath Kudlur, Josh Levenberg, Rajat Monga, Sherry Moore, Derek~G. Murray, Benoit Steiner, Paul Tucker, Vijay Vasudevan, Pete Warden, Martin Wicke, Yuan Yu, and Xiaoqiang Zheng.
\newblock {TensorFlow}: A system for {Large-Scale} machine learning.
\newblock In {\em 12th USENIX Symposium on Operating Systems Design and Implementation (OSDI 16)}, pages 265--283, Savannah, GA, November 2016. USENIX Association.

\bibitem{10.5555/3433701.3433792}
Ahsan Ali, Riccardo Pinciroli, Feng Yan, and Evgenia Smirni.
\newblock Batch: machine learning inference serving on serverless platforms with adaptive batching.
\newblock In {\em Proceedings of the International Conference for High Performance Computing, Networking, Storage and Analysis}, SC '20. IEEE Press, 2020.

\bibitem{amd2024rocm}
AMD.
\newblock {ROCm Examples}.
\newblock \url{https://github.com/ROCm/rocm-examples}, 2024.

\bibitem{ansel2009dmtcp}
Jason Ansel, Kapil Arya, and Gene Cooperman.
\newblock Dmtcp: Transparent checkpointing for cluster computations and the desktop.
\newblock In {\em 2009 IEEE International Symposium on Parallel \& Distributed Processing}, pages 1--12, 2009.

\bibitem{belotti2025memory}
Jonathon Belotti.
\newblock Memory snapshots: Checkpoint/restore for sub-second startup.
\newblock \url{https://modal.com/blog/mem-snapshots}, 2025.
\newblock Accessed: 2025-02-18.

\bibitem{chaudhary2020balancing}
Shubham Chaudhary, Ramachandran Ramjee, Muthian Sivathanu, Nipun Kwatra, and Srinidhi Viswanatha.
\newblock Balancing efficiency and fairness in heterogeneous gpu clusters for deep learning.
\newblock In {\em Proceedings of the Fifteenth European Conference on Computer Systems}, EuroSys '20, New York, NY, USA, 2020. Association for Computing Machinery.

\bibitem{chen2021evaluating}
Mark Chen, Jerry Tworek, Heewoo Jun, et~al.
\newblock Evaluating large language models trained on code.
\newblock \url{https://arxiv.org/abs/2107.03374}, 2021.

\bibitem{chen2015mxnet}
Tianqi Chen, Mu~Li, Yutian Li, Min Lin, Naiyan Wang, Minjie Wang, Tianjun Xiao, Bing Xu, Chiyuan Zhang, and Zheng Zhang.
\newblock Mxnet: A flexible and efficient machine learning library for heterogeneous distributed systems, 2015.

\bibitem{280768}
Seungbeom Choi, Sunho Lee, Yeonjae Kim, Jongse Park, Youngjin Kwon, and Jaehyuk Huh.
\newblock Serving heterogeneous machine learning models on {Multi-GPU} servers with {Spatio-Temporal} sharing.
\newblock In {\em 2022 USENIX Annual Technical Conference (USENIX ATC 22)}, pages 199--216, Carlsbad, CA, July 2022. USENIX Association.

\bibitem{cdi}
{Cloud Native Computing Foundation}.
\newblock Container device interface.
\newblock \url{https://github.com/cncf-tags/container-device-interface}.

\bibitem{corporation2023cuda}
NVIDIA Corporation.
\newblock Cuda c++ best practices guide.
\newblock \url{https://docs.nvidia.com/cuda/cuda-c-programming-guide/index.html#best-practices}, 2023.

\bibitem{criu-statistics}
{CRIU}.
\newblock {CRIU Statistics}.
\newblock \url{https://criu.org/Statistics}, 2024.

\bibitem{eiling2023cricket}
Niklas Eiling.
\newblock Cricket shared object support.
\newblock \url{https://github.com/RWTH-ACS/cricket/pull/15}, 2023.

\bibitem{eiling2022cricket}
Niklas Eiling, Jonas Baude, Stefan Lankes, and Antonello Monti.
\newblock Cricket: A virtualization layer for distributed execution of cuda applications with checkpoint/restart support.
\newblock {\em Concurrency and Computation: Practice and Experience}, 34(14):e6474, 2022.

\bibitem{eiling2023checkpoint}
Niklas Eiling, Stefan Lankes, and Antonello Monti.
\newblock Checkpoint/restart for cuda kernels.
\newblock In {\em Proceedings of the SC '23 Workshops of The International Conference on High Performance Computing, Network, Storage, and Analysis}, SC-W '23, page 1729–1737, New York, NY, USA, 2023. Association for Computing Machinery.

\bibitem{eisenman2022check}
Assaf Eisenman, Kiran~Kumar Matam, Steven Ingram, Dheevatsa Mudigere, Raghuraman Krishnamoorthi, Krishnakumar Nair, Misha Smelyanskiy, and Murali Annavaram.
\newblock {Check-N-Run}: a checkpointing system for training deep learning recommendation models.
\newblock In {\em 19th USENIX Symposium on Networked Systems Design and Implementation (NSDI 22)}, pages 929--943, Renton, WA, April 2022. USENIX Association.

\bibitem{linux-ptrace}
The~Linux Foundation.
\newblock {\em ptrace(2) - Linux manual page}, 2023.
\newblock Accessed: 2025-02-21.

\bibitem{goodfellow2016deep}
Ian Goodfellow, Yoshua Bengio, and Aaron Courville.
\newblock {\em Deep Learning}.
\newblock MIT Press, 2016.
\newblock \url{http://www.deeplearningbook.org}.

\bibitem{dubey2024llama}
Aaron Grattafiori, Abhimanyu Dubey, Abhinav Jauhri, et~al.
\newblock {The Llama 3 Herd of Models}, 2024.
\newblock arXiv:2407.21783 [cs.AI].

\bibitem{gupta2024just}
Tanmaey Gupta, Sanjeev Krishnan, Rituraj Kumar, Abhishek Vijeev, Bhargav Gulavani, Nipun Kwatra, Ramachandran Ramjee, and Muthian Sivathanu.
\newblock Just-in-time checkpointing: Low cost error recovery from deep learning training failures.
\newblock In {\em Proceedings of the Nineteenth European Conference on Computer Systems}, EuroSys '24, page 1110–1125, New York, NY, USA, 2024. Association for Computing Machinery.

\bibitem{gurfinkel2024checkpointing}
Steven Gurfinkel.
\newblock Checkpointing cuda applications with criu.
\newblock \url{https://developer.nvidia.com/blog/checkpointing-cuda-applications-with-criu/}, 2024.

\bibitem{hhargrove2006berkeley}
Paul~H Hargrove and Jason~C Duell.
\newblock Berkeley lab checkpoint/restart (blcr) for linux clusters.
\newblock {\em Journal of Physics: Conference Series}, 46(1):494, sep 2006.

\bibitem{harris2024cuda}
Mark Harris.
\newblock {Understand Fat Binaries and JIT Caching}.
\newblock \url{https://developer.nvidia.com/blog/cuda-pro-tip-understand-fat-binaries-jit-caching}, December 2013.

\bibitem{ishakian2018serving}
Vatche Ishakian, Vinod Muthusamy, and Aleksander Slominski.
\newblock Serving deep learning models in a serverless platform.
\newblock In {\em 2018 IEEE International Conference on Cloud Engineering (IC2E)}, pages 257--262. IEEE, 2018.

\bibitem{jain2020crac}
Twinkle Jain and Gene Cooperman.
\newblock Crac: Checkpoint-restart architecture for cuda with streams and uvm.
\newblock In {\em SC20: International Conference for High Performance Computing, Networking, Storage and Analysis}, pages 1--15, 2020.

\bibitem{kiran2022deep}
B~Ravi Kiran, Ibrahim Sobh, Victor Talpaert, Patrick Mannion, Ahmad A.~Al Sallab, Senthil Yogamani, and Patrick Pérez.
\newblock Deep reinforcement learning for autonomous driving: A survey.
\newblock {\em IEEE Transactions on Intelligent Transportation Systems}, 23(6):4909--4926, 2022.

\bibitem{280704}
Jie Li, Laiping Zhao, Yanan Yang, Kunlin Zhan, and Keqiu Li.
\newblock Tetris: Memory-efficient serverless inference through tensor sharing.
\newblock In {\em 2022 USENIX Annual Technical Conference (USENIX ATC 22)}, Carlsbad, CA, July 2022. USENIX Association.

\bibitem{ma2023dolphins}
Yingzi Ma, Yulong Cao, Jiachen Sun, Marco Pavone, and Chaowei Xiao.
\newblock Dolphins: Multimodal language model for driving, 2023.

\bibitem{microsoft2023github}
Microsoft.
\newblock Github copilot.
\newblock \url{https://copilot.microsoft.com/}, 2023.
\newblock Accessed: 2025-02-11.

\bibitem{copilot}
Microsoft.
\newblock {Microsoft 365 Copilot}.
\newblock \url{https://copilot.microsoft.com}, 2023.

\bibitem{microsoft2024boost}
Microsoft.
\newblock {Boost Checkpoint Speed and Reduce Cost with Nebula}.
\newblock \url{https://learn.microsoft.com/en-us/azure/machine-learning/reference-checkpoint-performance-for-large-models}, 2024.

\bibitem{mohan2021checkfreq}
Jayashree Mohan, Amar Phanishayee, and Vijay Chidambaram.
\newblock {CheckFreq}: Frequent, {Fine-Grained} {DNN} checkpointing.
\newblock In {\em 19th USENIX Conference on File and Storage Technologies (FAST 21)}, pages 203--216. USENIX Association, February 2021.

\bibitem{morgan2023ollama}
Jeffrey Morgan and Michael Chiang.
\newblock Ollama.
\newblock \url{https://ollama.ai/}, 2023.

\bibitem{moritz2018ray}
Philipp Moritz, Robert Nishihara, Stephanie Wang, Alexey Tumanov, Richard Liaw, Eric Liang, Melih Elibol, Zongheng Yang, William Paul, Michael~I. Jordan, and Ion Stoica.
\newblock Ray: A distributed framework for emerging {AI} applications.
\newblock In {\em 13th USENIX Symposium on Operating Systems Design and Implementation (OSDI 18)}, pages 561--577, Carlsbad, CA, October 2018. USENIX Association.

\bibitem{nicolae2020deepfreeze}
Bogdan Nicolae, Jiali Li, Justin~M. Wozniak, George Bosilca, Matthieu Dorier, and Franck Cappello.
\newblock Deepfreeze: Towards scalable asynchronous checkpointing of deep learning models.
\newblock In {\em 2020 20th IEEE/ACM International Symposium on Cluster, Cloud and Internet Computing (CCGRID)}, pages 172--181, 2020.

\bibitem{nukada2011nvcr}
Akira Nukada, Hiroyuki Takizawa, and Satoshi Matsuoka.
\newblock Nvcr: A transparent checkpoint-restart library for nvidia cuda.
\newblock In {\em 2011 IEEE International Symposium on Parallel and Distributed Processing Workshops and Phd Forum}, pages 104--113, 2011.

\bibitem{container-toolkit}
{NVIDIA}.
\newblock {Container Toolkit}.
\newblock \url{https://github.com/NVIDIA/nvidia-container-toolkit}.

\bibitem{nvidia2017p100}
NVIDIA.
\newblock {Tesla P100}.
\newblock White paper, NVIDIA, 2017.

\bibitem{nvidia2020a100}
NVIDIA.
\newblock {A100 Tensor Core GPU Architecture}.
\newblock White paper, NVIDIA, 2020.

\bibitem{nvidia2022h100}
NVIDIA.
\newblock {H100 Tensor Core GPU Architecture}.
\newblock White paper, NVIDIA, 2022.

\bibitem{nvidia2023introduction}
NVIDIA.
\newblock {Introduction to LLM Agents}.
\newblock \url{https://developer.nvidia.com/blog/introduction-to-llm-agents}, Nov 2023.

\bibitem{nvidia2024driver}
NVIDIA.
\newblock {Driver API Documentation}.
\newblock \url{https://docs.nvidia.com/cuda/cuda-driver-api/index.html}, 2024.
\newblock Accessed: 2024-12-01.

\bibitem{nvidia2024megatron-lm}
NVIDIA.
\newblock Megatron-lm.
\newblock \url{https://github.com/NVIDIA/Megatron-LM}, 2024.

\bibitem{nvidia2024runtime}
NVIDIA.
\newblock {Runtime API Documentation}.
\newblock \url{https://docs.nvidia.com/cuda/cuda-runtime-api/index.html}, 2024.
\newblock Accessed: 2024-12-01.

\bibitem{chatgpt}
Long Ouyang, Jeffrey Wu, Xu~Jiang, et~al.
\newblock Training language models to follow instructions with human feedback.
\newblock In S.~Koyejo, S.~Mohamed, A.~Agarwal, D.~Belgrave, K.~Cho, and A.~Oh, editors, {\em Advances in Neural Information Processing Systems}, volume~35, pages 27730--27744. Curran Associates, Inc., 2022.

\bibitem{park2022gpureplay}
Heejin Park and Felix~Xiaozhu Lin.
\newblock Gpureplay: a 50-kb gpu stack for client ml.
\newblock In {\em Proceedings of the 27th ACM International Conference on Architectural Support for Programming Languages and Operating Systems}, ASPLOS '22, page 157–170, New York, NY, USA, 2022. Association for Computing Machinery.

\bibitem{paszke2019pytorch}
Adam et~al. Paszke.
\newblock Pytorch: An imperative style, high-performance deep learning library.
\newblock In H.~Wallach, H.~Larochelle, A.~Beygelzimer, F.~d\textquotesingle Alch\'{e}-Buc, E.~Fox, and R.~Garnett, editors, {\em Advances in Neural Information Processing Systems}, volume~32. Curran Associates, Inc., 2019.

\bibitem{10.1145/3620678.3624664}
Qiangyu Pei, Yongjie Yuan, Haichuan Hu, Qiong Chen, and Fangming Liu.
\newblock Asyfunc: A high-performance and resource-efficient serverless inference system via asymmetric functions.
\newblock In {\em Proceedings of the 2023 ACM Symposium on Cloud Computing}, SoCC '23, page 324–340, New York, NY, USA, 2023. Association for Computing Machinery.

\bibitem{rajbhandari2020zero}
Samyam Rajbhandari, Jeff Rasley, Olatunji Ruwase, and Yuxiong He.
\newblock Zero: Memory optimizations toward training trillion parameter models.
\newblock In {\em SC20: International Conference for High Performance Computing, Networking, Storage and Analysis}, pages 1--16. IEEE, 2020.

\bibitem{bhardwaj2021drm}
{Rajneesh Bhardwaj}.
\newblock {drm/amdkfd: CRIU Introduce Checkpoint-Restore APIs}, 2021.
\newblock Linux Commit ID: \href{https://git.kernel.org/pub/scm/linux/kernel/git/torvalds/linux.git/commit/?id=36988070}{3698807094ecae945436921325f5c309d1123f11}.

\bibitem{bhardwaj2021fast}
David Yat~Sin Rajneesh~Bhardwaj, Felix~Kuehling.
\newblock Fast checkpoint restore for gpus.
\newblock \url{https://lpc.events/event/11/contributions/891}, 2021.
\newblock Accessed: 2023-08-15.

\bibitem{riachnvidia2019gtc}
Duncan Riach.
\newblock Determinism in deep learning.
\newblock In {\em GTC Silicon Valley}. NVIDIA, 2019.

\bibitem{richards2015university}
Andrew Richards.
\newblock {\em University of Oxford Advanced Research Computing}, August 2015.

\bibitem{rojas2019analyzing}
Elvis Rojas, Esteban Meneses, Terry Jones, and Don Maxwell.
\newblock Analyzing a five-year failure record of a leadership-class supercomputer.
\newblock In {\em 2019 31st International Symposium on Computer Architecture and High Performance Computing (SBAC-PAD)}, pages 196--203, 2019.

\bibitem{infaas}
Francisco Romero, Qian Li, Neeraja~J. Yadwadkar, and Christos Kozyrakis.
\newblock {INFaaS}: Automated model-less inference serving.
\newblock In {\em 2021 USENIX Annual Technical Conference (USENIX ATC 21)}, pages 397--411. USENIX Association, July 2021.

\bibitem{roziere2024code}
Baptiste Rozière, Jonas Gehring, Fabian Gloeckle, Sten Sootla, Itai Gat, Xiaoqing~Ellen Tan, Yossi Adi, Jingyu Liu, Romain Sauvestre, Tal Remez, Jérémy Rapin, Artyom Kozhevnikov, Ivan Evtimov, Joanna Bitton, Manish Bhatt, Cristian~Canton Ferrer, Aaron Grattafiori, Wenhan Xiong, Alexandre Défossez, Jade Copet, Faisal Azhar, Hugo Touvron, Louis Martin, Nicolas Usunier, Thomas Scialom, and Gabriel Synnaeve.
\newblock Code llama: Open foundation models for code, 2024.

\bibitem{shukla2022singularity}
Dharma Shukla, Muthian Sivathanu, Srinidhi Viswanatha, Bhargav Gulavani, Rimma Nehme, Amey Agrawal, Chen Chen, Nipun Kwatra, Ramachandran Ramjee, Pankaj Sharma, Atul Katiyar, Vipul Modi, Vaibhav Sharma, Abhishek Singh, Shreshth Singhal, Kaustubh Welankar, Lu~Xun, Ravi Anupindi, Karthik Elangovan, Hasibur Rahman, Zhou Lin, Rahul Seetharaman, Cheng Xu, Eddie Ailijiang, Suresh Krishnappa, and Mark Russinovich.
\newblock Singularity: Planet-scale, preemptive and elastic scheduling of ai workloads, 2022.

\bibitem{sivathanu2022transparent}
Muthian Sivathanu, Srinidhi Viswanatha, Dharma~Kiritkumar Shukla, Nipun Kwatra, Ramachandran Ramjee, Rimma~Vladimirovna Nehme, Pankaj Sharma, Bhalakumaaran~Erode Ranganathan, and Vaibhav Sharma.
\newblock Transparent pre-emption and migration for planet-scale computer, September 2022.
\newblock US Patent App. 17/359,553.

\bibitem{cuda-checkpoint}
{Steven Gurfinkel}.
\newblock {CUDA Checkpoint and Restore Utility}.
\newblock \url{https://github.com/NVIDIA/cuda-checkpoint}.

\bibitem{takizawa2009checuda}
Hiroyuki Takizawa, Katsuto Sato, Kazuhiko Komatsu, and Hiroaki Kobayashi.
\newblock Checuda: A checkpoint/restart tool for cuda applications.
\newblock In {\em 2009 International Conference on Parallel and Distributed Computing, Applications and Technologies}, pages 408--413, 2009.

\bibitem{criu}
CRIU Team.
\newblock {Checkpoint/Restore In Userspace}.
\newblock \url{https://criu.org/}.

\bibitem{team2024gemini}
Gemini Team.
\newblock Gemini: {A} {Family} of {Highly} {Capable} {Multimodal} {Models}, 2024.
\newblock arXiv:2312.11805 [cs.CL].

\bibitem{wang2023gemini}
Zhuang Wang, Zhen Jia, Shuai Zheng, Zhen Zhang, Xinwei Fu, T.~S.~Eugene Ng, and Yida Wang.
\newblock Gemini: Fast failure recovery in distributed training with in-memory checkpoints.
\newblock In {\em Proceedings of the 29th Symposium on Operating Systems Principles}, SOSP '23, page 364–381, New York, NY, USA, 2023. Association for Computing Machinery.

\bibitem{wu2025memory}
Bernie Wu.
\newblock Memverge memory machine ai gpu-as-a-service.
\newblock \url{https://techfieldday.com/video/memverge-memory-machine-ai-gpu-as-a-service/}, 2025.
\newblock Accessed: 2025-02-18.

\bibitem{10.1145/3503222.3507709}
Yanan Yang, Laiping Zhao, Yiming Li, Huanyu Zhang, Jie Li, Mingyang Zhao, Xingzhen Chen, and Keqiu Li.
\newblock Infless: a native serverless system for low-latency, high-throughput inference.
\newblock In {\em Proceedings of the 27th ACM International Conference on Architectural Support for Programming Languages and Operating Systems}, ASPLOS '22, page 768–781, New York, NY, USA, 2022. Association for Computing Machinery.

\bibitem{yang2024on-demand}
Yanning Yang, Dong Du, Haitao Song, and Yubin Xia.
\newblock On-demand and parallel checkpoint/restore for gpu applications.
\newblock In {\em Proceedings of the 2024 ACM Symposium on Cloud Computing}, SoCC '24, page 415–433, New York, NY, USA, 2024. Association for Computing Machinery.

\bibitem{yang2024part}
Zeyu Yang, Karel Adamek, and Wesley Armour.
\newblock Part-time power measurements: nvidia-smi's lack of attention, 2024.

\bibitem{234998}
Chengliang Zhang, Minchen Yu, Wei Wang, and Feng Yan.
\newblock {MArk}: Exploiting cloud services for {Cost-Effective}, {SLO-Aware} machine learning inference serving.
\newblock In {\em 2019 USENIX Annual Technical Conference (USENIX ATC 19)}, pages 1049--1062, Renton, WA, July 2019. USENIX Association.

\end{thebibliography}

\end{document}